\documentclass[]{aastex62}

\newcommand{\italicbold}[1]{\textbf{\textit{#1}}}
\newcommand{\y}[1]{{\textcolor{red}{#1}}}

\graphicspath{{./}{figures/}}
\shortauthors{Seo et al.}
\usepackage{lipsum}
\usepackage{float}
\usepackage{kotex}
\usepackage{tabularx}
\usepackage{mathtools}
\usepackage{verbatim}

\begin{document}

\title{A New Rarity Assessment of the `Disk of Satellites': the Milky Way System Is the Exception Rather than the Rule \\ in the $\Lambda$CDM Cosmology}

\correspondingauthor{Suk-Jin Yoon}
\email{sjyoon0691@yonsei.ac.kr}

\author[0000-0002-0749-2090]{Chanoul Seo}
\affil{Equal First Authors}
\affiliation{Department of Astronomy, Yonsei University, Seoul 03722, Republic of Korea}
\affiliation{Department of Astronomical Science, Graduate University for Advanced Studies (SOKENDAI), Tokyo 181-8588, Japan}
\affiliation{National Astronomical Observatory of Japan, Tokyo 181-8588, Japan}

\author[0000-0002-1842-4325]{Suk-Jin Yoon}
\affil{Equal First Authors}
\affiliation{Department of Astronomy, Yonsei University, Seoul 03722, Republic of Korea}
\affiliation{Center for Galaxy Evolution Research, Yonsei University, Seoul 03722, Republic of Korea}

\author[0000-0003-2922-6866]{Sanjaya Paudel}
\affiliation{Department of Astronomy, Yonsei University, Seoul 03722, Republic of Korea}
\affiliation{Center for Galaxy Evolution Research, Yonsei University, Seoul 03722, Republic of Korea}

\author[0000-0003-3791-0860]{Sung-Ho An}
\affiliation{Department of Earth Science Education, Daegu University, Gyeongsangbuk-do 38453, Republic of Korea}

\author[0000-0001-7075-4156]{Jun-Sung Moon}
\affiliation{Astronomy Program, Department of Physics and Astronomy, Seoul National University, Seoul 08826, Republic of Korea}
\affiliation{Research Institute of Basic Sciences, Seoul National University, Seoul 08826, Republic of Korea}

\begin{abstract}
The majority of satellite galaxies around the Milky Way (MW) show disk-like distributions (the disk of satellites; DoS), which is a small-scale problem of the $\Lambda$CDM cosmology. 
The conventional definition of the MW-like DoS is a satellite system with a minor-to-major axis ratio ($c$/$a$) lower than the MW's $c$/$a$ value of 0.181. 
Here we question the validity of the $c$/$a$-based DoS rarity assessment and propose an alternative approach. 
How satellites are placed around a galaxy is dictated mainly by two factors: the distributions of satellites’ orbital poles and distances from the host. 
Based on this premise, we construct the `satellite distribution generator' code and generate 10$^5$ `spatially and kinematically analogous systems (SKASs)' sharing these two factors. 
The SKAS can disclose the intrinsic, underlying $c$/$a$ probability distribution function (PDF), from which a present-day $c$/$a$ value is fortuitously determined.
We find that the $c$/$a$ PDF of the MW DoS defined by 11 classical satellites
is quite broad ($\sigma_{c/a}$\,$\sim$\,0.105), implying that a simple present-day $c$/$a$ value, combined with its highly time-variable nature, cannot fully represent the degree of flatness. 
Moreover, based on the intrinsic $c$/$a$ PDF, we re-evaluate the rarity of the MW DoS by comparing it with IllustrisTNG50-1 host--satellite systems and find that even with the new measure, the MW DoS remains {\it rare} (0.00$\sim$3.40\%). 
We show that the reason behind the rareness is that {\it both} orbital poles and distances of the 11 MW satellites are far more {\it plane-friendly} than those of simulated host--satellite systems, challenging the current structure and galaxy formation model.
\end{abstract}
\keywords{Dwarf galaxies (416) --- Galaxy kinematics (602) --- Milky Way dynamics (1051)}

\section{Introduction}
\label{section:intro}

The satellite galaxies around the Milky Way (MW) show a flattened, disk-like distribution. Current Lambda Cold Dark Matter ($\Lambda$CDM) cosmological simulations, however, indicate that this type of distribution is highly uncommon. This discrepancy poses one of the smaller-scale challenges of the $\Lambda$CDM cosmology. The anisotropic distribution of satellite galaxies around the MW was initially observed by \citet{1976MNRAS.174..695L} and \citet{1976RGOB..182..241K}. Subsequent investigations have consistently verified the existence of this flattened structure \cite[e.g.,][]{2007MNRAS.374.1125M, 2009MNRAS.394.2223M, 2010A&A...523A..32K, 2012MNRAS.423.1109P, 2014ApJ...790...74P, 2015MNRAS.453.1047P} and the alignment of orbital poles among the satellites \cite[e.g.,][]{2008ApJ...680..287M, 2013MNRAS.435.2116P, 2018A&A...619A.103F, 2019MNRAS.488.1166S, 2020MNRAS.491.3042P, 2021ApJ...916....8L}. This structural configuration, formed by the satellite galaxies encircling the MW, is commonly denoted as the `disk of satellites (DoS; \citet{2008ApJ...680..287M})', or as the `Vast Polar Structure (VPOS; \citet{2012MNRAS.423.1109P})' when encompassing additional substructures such as globular clusters \cite[e.g.,][]{2002Sci...297..578Y} and halo stellar streams.

The identification of the DoS around the MW has prompted a surge in research exploring the presence of analogous anisotropic and rotationally supported structures in other galaxies. Notable investigations include those focused on M31 \cite[e.g.,][]{2007MNRAS.374.1125M, 2009MNRAS.394.2223M, 2013Natur.493...62I, 2013ApJ...766..120C, 2020ApJ...901...43S}, NGC 3109 \cite[]{2013A&A...559L..11B}, and systems beyond the Local Group such as Centaurus A \cite[e.g.,][]{2015ApJ...802L..25T, 2015A&A...583A..79M, 2016A&A...595A.119M, 2019A&A...629A..18M}, SDSS galaxy pairs (e.g., \citealt{2014Natur.511..563I, 2015ApJ...805...67I, 2024NatAs.tmp...26G}; see \citealt{2015MNRAS.453.3839P} and \citealt{2015MNRAS.449.2576C} for related discussions), the M101 group complex \cite[]{2017A&A...602A.119M}, M81 \cite[]{2013AJ....146..126C}, NGC 2750 \cite[]{2021ApJ...917L..18P}, NGC 253 (\citealt{2021A&A...652A..48M}; see \citealt{2024arXiv240114457M} for the ongoing debate), and massive host galaxies in the MATLAS survey \cite[]{2021A&A...654A.161H}. Moreover, proposals have been made regarding the existence of large thin structures in the Local Group as a whole \cite[e.g.,][]{2013MNRAS.435.1928P, 2014MNRAS.440..908P}.

The proposition by \citet{2005A&A...431..517K} that the flattened structure around the MW poses a challenge to the $\Lambda$CDM cosmology initiated debates. Subsequent studies, however, raised questions about this hypothesis and suggested that the MW DoS may be consistent with the $\Lambda$CDM cosmology \cite[e.g.,][]{2005A&A...437..383K, 2005ApJ...629..219Z, 2005MNRAS.363..146L, 2007MNRAS.374...16L, 2009MNRAS.399..550L}. Subsequent studies have incorporated various updates, such as considerations of kinematic coherence, an increased number of plane members, new plane-finding algorithms, observational coverage, and the inclusion of baryonic physics \cite[e.g.,][]{2008ApJ...680..287M, 2011MNRAS.415.2607D, 2012MNRAS.424...80P, 2013MNRAS.429.1502W, 2014arXiv1412.2748S, 2015ApJ...815...19P, 2015MNRAS.452.3838C, 2016MNRAS.457.1931S, 2017MNRAS.466.3119A, 2018MNRAS.478.5533F, 2019MNRAS.488.1166S, 2019ApJ...875..105P, 2020ApJ...897...71S, 2020MNRAS.499.3755S, 2020MNRAS.491.3042P, 2021MNRAS.504.1379S, 2022MNRAS.514..390G, 2023NatAs...7..481S, 2023ApJ...954..128X, 2023Galax..11..114Z}. The discourse on the statistical significance of the planar alignments of satellites has expanded to include considerations of the rarity of planes in the $\Lambda$CDM cosmology for other galaxies \cite[e.g.,][]{2014ApJ...784L...6I, 2015ApJ...800...34G, 2015MNRAS.452.3838C, 2016MNRAS.460.4348B, 2017MNRAS.466.3119A, 2018Sci...359..534M, 2018MNRAS.478.5533F, 2020MNRAS.499.3755S, 2021MNRAS.504.1379S, 2021ApJ...923...42P}.

Various potential cosmological origins for the anisotropic distribution of satellites have been proposed. These include the hypothesis of tidal dwarf galaxies \cite[e.g.,][]{2011A&A...532A.118P, 2012MNRAS.424...80P, 2012MNRAS.423.1109P, 2013MNRAS.435.2116P}, filamentary accretion \cite[e.g.,][]{2005ApJ...629..219Z, 2009ApJ...697..269M, 2011MNRAS.413.3013L, 2017MNRAS.466.3119A, 2018A&A...613A...4W, 2018MNRAS.476.1796S, 2020ApJ...900..129W}, group infall (e.g., \citealt{2008MNRAS.385.1365L, 2008ApJ...686L..61D, 2009ApJ...697..269M, 2013MNRAS.429.1502W, 2018MNRAS.476.1796S, 2024A&A...681A..73T}; see \citealt{2021MNRAS.504.4551S}, \citealt{2021ApJ...923..140G}, and \citealt{2022ApJ...932...70P} for the discussions on the LMC's effect). A comprehensive overview of the planar alignment problem is available in the review papers \cite[]{2018MPLA...3330004P, 2021Galax...9...66P} and the associated commentary \cite[]{2021NatAs...5.1185P}.

A conventional criterion for expressing the flatness of a planar alignment structure is the observed minor-to-major axis ratio ($c$/$a$) of a satellite distribution \cite[e.g.,][]{2005MNRAS.363..146L, 2014MNRAS.442.2362P, 2016MNRAS.457.1931S, 2019MNRAS.488.1166S, 2022MNRAS.514..390G}, along with the absolute plane height \cite[e.g.,][]{2005A&A...437..383K, 2014MNRAS.442.2362P, 2020MNRAS.491.3042P}. A lower $c$/$a$ indicates a more flattened structure. The conventional definition of a `MW-like DoS' is one whose currently observed $c$/$a$ is equal to or lower than that of the MW. In the present study, we challenge the conventional measure of flatness, which relies solely on $c$/$a$ at a specific moment, and propose an alternative method for quantifying the intrinsic flatness of a host--satellites system. Leveraging the understanding that satellites' orbital pole and radial distribution are crucial parameters influencing flatness \cite[e.g.,][]{2005ApJ...629..219Z, 2014MNRAS.442.2362P, 2015MNRAS.452.3838C, 2017MNRAS.466.3119A, 2022MNRAS.514..390G}, we introduce the concept of `intrinsic $c$/$a$ probability distribution function (PDF),' which provides a new way to examine the flatness of a system. Applying this concept, we reevaluate the rarity of the MW DoS and re-assess the planar alignment problem. This study focuses specifically on the conventional MW DoS, comprising the classical 11 satellites \cite[e.g.,][]{2019MNRAS.488.1166S, 2022MNRAS.514..390G}.

This paper is organized as follows. In Section \ref{sec2}, detailed information is provided regarding the observation and cosmological simulation data sets employed in this study. Section \ref{sec3} outlines the identification of a host--satellite system's DoS in the IllustrisTNG simulation (the highest mass-resolution TNG50-1 suite), calculates the conventional $c$/$a$ values, and estimates the rarity of the MW DoS by comparing its $c$/$a$ with those of systems in the simulation.
In Section \ref{sec4}, we construct the `satellite distribution generator' code and generate `spatially and kinematically analogous systems (SKASs)' sharing the distributions of satellites’ orbital poles and distances from the host. The SKASs can disclose the intrinsic, underlying $c$/$a$ PDF. Based on the $c$/$a$ PDF, we re-evaluate the rarity of the MW DoS by comparing it with IllustrisTNG50-1 and find that even with the new measure, the MW DoS remains quite rare (2.48\%) in the $\Lambda$CDM cosmology. 
Section \ref{sec5} delves into the underlying reasons for the rareness of the MW DoS by parameterizing two flatness-related factors and exploring their relationship with the $c$/$a$ PDF. In Section \ref{sec6}, we discuss the effects of different satellite selection criteria, the increased tension between the MW and simulation when the orbital direction is considered, and the potential impact of astrophysical effects on our findings. Finally, we summarize and conclude our results in Section \ref{sec7}.

\section{OBSERVATIONAL AND SIMULATION DATA} \label{sec2}

\subsection{Observational Data}
\label{ss:ObservationalData}

In this section, we introduce the MW observational data for the 11 classical satellites used in this study.
Table \ref{tab:used data} summarizes the position and velocity data for the MW and its 11 classical satellites, along with their total mass data. The total mass data for the 11 classical satellites are compiled from various sources, including \citet{2018MNRAS.481.5073E} (excluding the Magellanic Clouds), \citet{2021ApJ...923..149S} (pertaining to the LMC), and \citet{2004ApJ...604..176S} (pertaining to the SMC). The position and velocity data of the 11 satellites come from multiple studies as detailed below:

LMC, SMC, and Sagittarius dwarf spheroidal galaxy (Sgr dSph): We rely on the Local Group galaxy data provided by \citet{2020MNRAS.491.3042P}, which references \citet{2012AJ....144....4M} for sky coordinates and utilizes proper motion data from Gaia DR2 \cite[]{2018A&A...616A..12G} along with the best available pre-Gaia proper motion data \citep[and references therein]{2020MNRAS.491.3042P}.

Other classical satellites (except for Leo I and Leo II): We use the position and velocity data from Table 1 of \citet{2024A&A...681A..73T}, which refers to the celestial coordinates from \citet{2021ApJ...916....8L} and orbital parameters from \citet{2022A&A...657A..54B}, based on Gaia EDR3 \cite[]{2021A&A...649A...1G} data.

Leo I and Leo II: We refer to Table 5 of \citet{2024ApJ...971...98B}, which has lower uncertainties for Leo I and Leo II compared to \citet{2022A&A...657A..54B}. \citet{2024ApJ...971...98B} used distances and line-of-sight velocity data from \citet{2012AJ....144....4M} and their proper motion evaluation (GAIAHUB) using HST and Gaia EDR3 \cite[]{2021A&A...649A...1G} data.

\subsection{$\Lambda$CDM Cosmological Simulation Data}
\label{ss:Simulation}

In this study, we use the IllistrisTNG simulation \cite[]{2018MNRAS.475..648P,2018MNRAS.475..676S,2019ComAC...6....2N,2019MNRAS.490.3234N,2019MNRAS.490.3196P}. The IllustrisTNG project offers simulations in three different volumes (TNG50, TNG100, and TNG300). For our analysis, we specifically utilize the TNG50-1, which is the highest mass-resolution run. 
TNG50-1 features DM and baryonic particle masses of $m_{\text{DM}}=4.5 \times 10^5M_{\odot}$ and $m_{\text{gas}}=8.5 \times 10^4M_{\odot}$. 
The box size of the TNG50-1 simulation is 35\,$h^{-1}$Mpc.
The cosmological parameters used in the IllustrisTNG simulations are 0.3089 for matter density ($\Omega_{m}$), 0.0486 for baryonic density ($\Omega_{b}$), 0.6911 for cosmological constant ($\Omega_{\Lambda}$), 0.6774 for Hubble parameter ($h$), 0.8159 for density perturbation amplitude ($\sigma_{8}$), and 0.9667 for primordial spectral index ($n_{s}$). All these parameters are consistent with the Planck Collaboration XIII results (\citealt{2016A&A...594A..13P}).

The IllistrisTNG simulation provides the friends-of-friends (FoF) halo group and subhalo data based on the SUBFIND algorithm (\citealt{2001MNRAS.328..726S}). In our analysis, we define `MW-mass halos' as luminous subhalos ($M_{*} > 0$) with total masses within the range $[0.3 \times 10^{12}M_{\odot}, 3 \times 10^{12}M_{\odot}]$, accounting for uncertainties in MW mass estimations \cite[e.g.,][]{2019MNRAS.484.5453C, 2019MNRAS.488.1166S}. We refer to the most massive subhalos in the FoF halos as the main halos. Within the FoF halo groups, we consider subhalos with total masses less than that of the MW-mass main halo as satellite candidates. We further filter these satellite candidates using three criteria. First, we only consider luminous subhalos ($M_{*} > 0$) with a cosmological origin. Second, we apply a low-mass cut of $10^{8.5}M_{\odot}$ based on the masses of the 11 luminous classical satellites of the MW. 
Third, satellites should either be bound to the MW-mass main halo or undergo strong interactions that would eventually lead to mergers.

In this study, we consider the possibility that some main halo--satellite subhalo pairs within the same FoF halo group may not be bound \cite[e.g.,][]{2013MNRAS.436.1765M}. Instead using all subhalos identified by the SUBFIND algorithm, we apply additional criteria to identify subhalos that are bound to the host or sufficiently strongly interacting to eventually merge. In particular, we employ the total energy of the main halo--satellite pair to determine the binding status of the satellite to the MW-mass halo, following established methods \cite[e.g.,][]{2012MNRAS.425.2313T, 2013MNRAS.436.1765M, 2019ApJ...887...59A}. Utilizing the simple total energy equation of a two-body system \cite[]{2019ApJ...887...59A}, the total energy ($E_{12}$) can be calculated as 
\begin{equation}
E_{12} = M_{1}M_{2}\left\{\frac{|\boldsymbol{V_{1}}-\boldsymbol{V_{2}}+H(z) \boldsymbol{R_{12}}|^{2}}{2(M_{1}+M_{2})} - \frac{G}{| \boldsymbol{R_{12}}|}\right\},
\end{equation} 
where $M_{1}$ and $M_{2}$ respectively represent the masses of the main halo and satellite galaxies, $\boldsymbol{V_{1}}$ and $\boldsymbol{V_{2}}$ respectively denote the peculiar velocity vectors of the main halo and satellite galaxies with respect to the mutual center of mass, $H(z)$ is the Hubble parameter, and $\boldsymbol{R_{12}}$ is the displacement vector. A negative total energy indicates a bound system. However, it is important to note that dynamical friction can alter the total energy from positive to negative values, suggesting that an initially unbound (fly-by) system may merge over time \cite[e.g.,][]{2012ApJ...751...17S, 2019ApJ...887...59A}. In this regard, we adopt the assumption that a satellite undergoes a merger even when the total energy is positive, provided that the total energy is smaller than the capture criterion \cite[e.g.,][]{2003ApJ...582..141G, 2019ApJ...887...59A}. The capture criterion is defined as 
\begin{equation} 
\Delta E = 2.4\mu \frac{M_{2}}{M_{1}} \frac{\sigma_{1} ^{4}}{V_{\text{rel}}^{2} + \sigma_{1}^{2}},
\end{equation} 
where $\mu$ is the reduced mass, $M_{1}$ and $M_{2}$ are respectively the masses of the main and satellite galaxies, $\sigma_{1}$ is the velocity dispersion of the main halo, and $V_{\text{rel}}$ is the relative velocity. 
In Section \ref{ss:sampleselect}, we discuss the impact of two additional selection criteria. Our analysis shows that different satellite selection criteria introduce only minor differences in the number of selected subhalos and overall results, and do not affect our conclusions.

We define a FoF halo group as a `MW-analogous host--satellite system' if it satisfies the following conditions: ($a$) The MW-mass halo exists as the most massive subhalo; ($b$) The total mass ratio between the second most massive subhalo is smaller than 0.1 (approximately the LMC-MW mass ratio); and ($c$) The FoF group should contain at least 11 satellites. We specifically use the 11 brightest ($r$-band magnitude) satellites in the study to match the satellite number with the classical 11 satellites of the MW. A total of 202 groups are identified in the IllustrisTNG50-1 simulation as MW-analogous host--satellite systems. In Section \ref{ss:LMCmasseffect}, we discuss the effect of using different second most massive subhalo criteria.

\section{How Rare is the MW D\lowercase{o}S Based on the Conventional \lowercase{$c$/$a$} Values?}  \label{sec3}

\subsection{Plane Identification}
\label{ss:Plane Identification}

In this study, we employ the principal component analysis (PCA) method to quantify the shape of the spatial distribution of satellite galaxies. PCA provides information about the three main axes of the ellipsoid that best fits the three-dimensional distribution of satellites. The covariance matrix is calculated from the three-dimensional positional data of the 11 satellites using the following formula:
\begin{equation}
COV(i,j)=\frac{1}{N} \sum_{k=1}^{N}(i_{k}-m_{i})(j_{k}-m_{j}), 
\end{equation}
where $N$ is the number of satellite galaxies, $i$ and $j$ are the three components of the three-dimensional relative position vector ($x,y,z$ from the MW at the origin), and $m$ is the average value.

The three derived eigenvalues \{$\lambda_{1}$, $\lambda_{2}$, $\lambda_{3}$\} obtained through Eigen-Decomposition are arranged in descending order, and the corresponding eigenvectors \{$\italicbold{e}_{1}$, $\italicbold{e}_{2}$, $\italicbold{e}_{3}$\} represent the main axes of the ellipsoid. The ratio of the minor to major axis, $c$/$a$, is adopted as a measure of the flatness of the system. For the observed MW DoS, the $c$/$a$ ratio is found to be 0.181. This is slightly smaller than a previous study (0.183, from \citet{2019MNRAS.488.1166S}) that utilized the position data from \citet{2012AJ....144....4M}.

\subsection{The Rarity of the MW DoS Based on the Conventional $c$/$a$ Value}
\label{ss:MWraritywithpresentca}

\begin{figure}
\includegraphics[width=9cm]{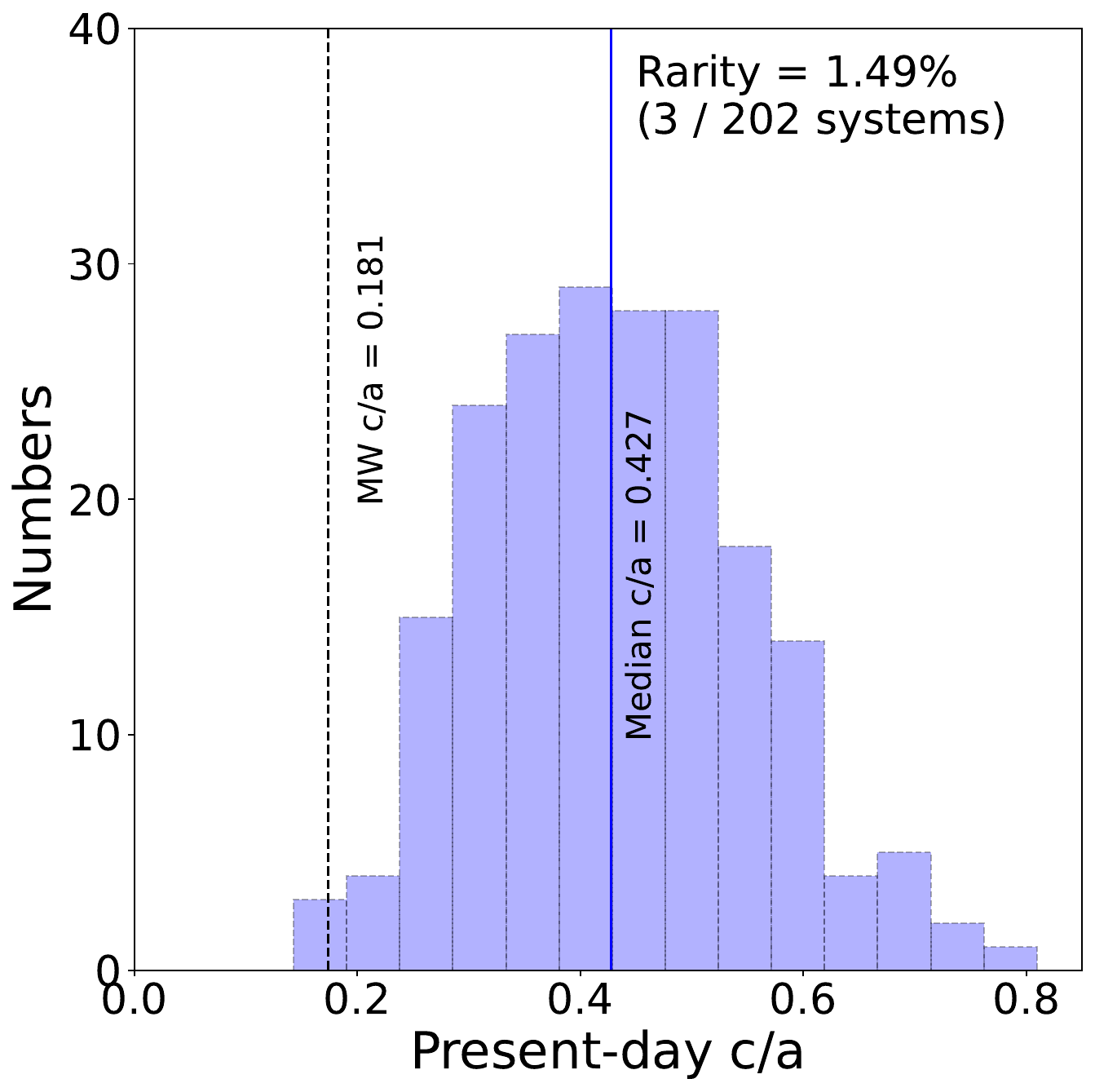}
\centering
\caption{
The $c$/$a$ distributions of 202 MW-analogous host--satellite systems in the TNG50-1 simulation. 
The x-axis represents present-day $c$/$a$ and the y-axis indicates the number of systems within each $c$/$a$ bin. 
The vertical solid blue line indicates the median $c$/$a$ of the total 202 systems, and the vertical dotted black line corresponds to the observed $c$/$a$ of the MW DoS. 
The degree of rarity of the MW DoS is 1.49\,\%.} 
\label{fig:presentca}
\end{figure}

Figure \ref{fig:presentca} illustrates the present-day $c$/$a$ distribution of 202 MW-analogous host--satellite systems in the Illustris TNG50-1 simulation. The median $c$/$a$ for the entire system is 0.427. If a system has a lower $c$/$a$ than that of the MW (= 0.181), it is considered `MW-like DoS,' comprising 1.49\,\% (3 out of 202 systems) of the total. This reaffirms the well-known rareness of the MW DoS in cosmological simulations.

For the rarity of the MW DoS, previous studies based on $c$/$a$ of the brightest (most massive) 11 satellites reported slightly lower (e.g., 0.75\,\% by \citealt{2020MNRAS.491.3042P} using the TNG100-1 simulation; 1\,\% by \citealt{2019MNRAS.488.1166S} using the EAGLE simulation) or higher values (e.g., 3--5\,\% by \citet{2014MNRAS.442.2362P} and \citet{2022MNRAS.514..390G} using the Millennium-II simulation) compared to ours (1.49\,\%). The difference among the studies likely arises from the differing choices of mass resolution of simulations (\citealt{2020MNRAS.491.3042P}). In addition, the discrepancy can be attributed to differences in the satellite selection criterion. Our sample selection method based on the total energy does not impose a strict radial distance limitation, potentially leading to a lower radial concentration of satellites compared to the previous studies (see also Section \ref{ss:sampleselect} for a discussion on results using the distance-limited satellite selection method). The influence of the degree of radial concentration is explored in Section \ref{sec5}.

\section{How Rare is the MW D\lowercase{o}S Based on the Intrinsic, Underlying \lowercase{$c$/$a$} Probability Distribution Function?} \label{sec4}

\subsection{The `Satellite Distribution Generator' and `Spatially and Kinematically Analogous Systems'}
\label{ss:KAS}

The positions of 11 satellites are uniquely defined by three parameters: the orbital pole vector (referred to as $\vec{l}$), radial distance (referred to as \textit{d}), and orbital phase angle (referred to as $\phi$). Among these parameters, the combination of the first two parameters of 11 satellites, \italicbold{L} = \{$\vec{l}_{1},\ \vec{l}_{2},\ \vec{l}_{3},\ ...\ \vec{l}_{11}$\} and \italicbold{D} = \{$d_{1},\ d_{2},\ d_{3},\ ...\ d_{11}$\}, are directly linked to the configuration of the host--satellite system. The \italicbold{L} and \italicbold{D} dictate the `alignment of orbital poles' and the `radial distribution of satellites, respectively,' both of which play pivotal roles in determining the flatness of a host--satellite system.

How the alignment of orbital poles relates to the system's flatness was investigated by \citet{2019MNRAS.488.1166S}. They identified systems in the EAGLE simulation that have satellites with orbital poles similar to the MW DoS, termed `MW-like orbit systems.' Their analysis showed that $\sim$\,30\,\% of the MW-like orbit systems have MW-like thin planes. The limited effect of the orbital pole alignment was attributed to the outliers with quite different orbital poles. 
On the other hand, previous studies have underscored the importance of the radial concentration of satellites in understanding the structure of host--satellite systems \citep{2005ApJ...629..219Z, 2008MNRAS.385.1365L, 2014MNRAS.442.2362P, 2015MNRAS.452.3838C, 2017MNRAS.466.3119A, 2022MNRAS.514..390G}. Systems with a higher radial concentration of satellites tend to exhibit more flattened structures than systems, where satellites are arranged more uniformly across the system.

\begin{figure*}
\centering \includegraphics[width=15cm]{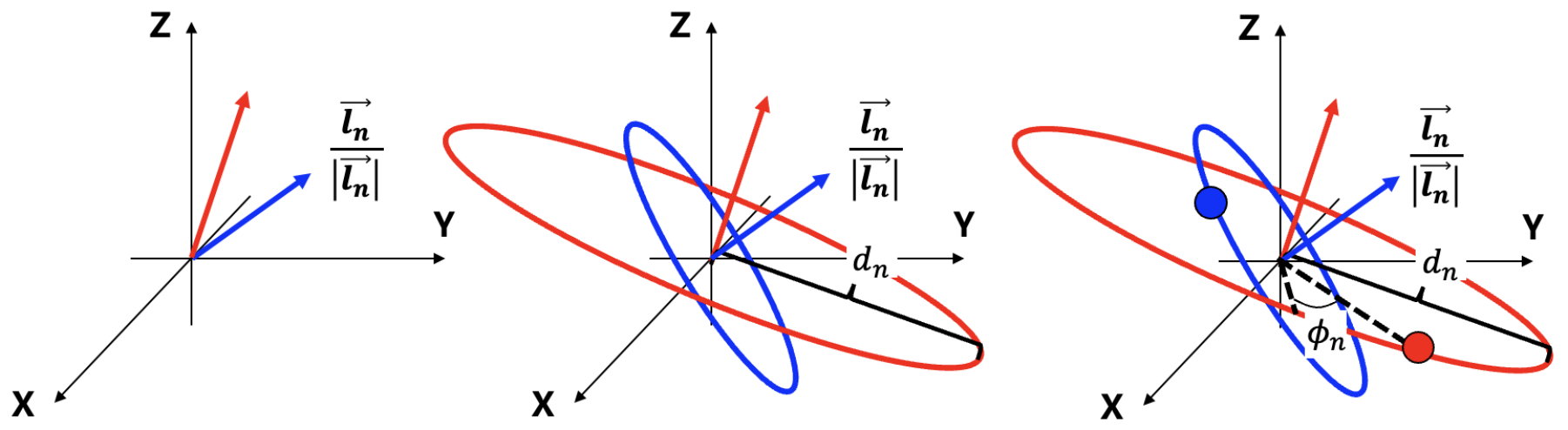}
\caption{
The process of constructing our satellite distribution generator (SDG) and generating the spatially and kinematically analogous systems (SKASs). 
({\it Left}) The orbital pole directions of satellites (\italicbold{L}) are initially set based on a simulated host--satellite system of interest. 
({\it Middle}) The distances from the host galaxy to the satellites (\italicbold{D}) are assigned based on the reference simulated system, forming circles along which satellites can be positioned. 
({\it Right}) The orbital phase angles ($\phi$) of satellites are given randomly on the circles, completing a system that meets the definition of the SKAS for the reference system.}
\label{fig:KASmechanism}
\end{figure*}

Our idea originates from the concept that how satellites are distributed around a galaxy is dictated by two factors: satellites' angular momentum directions and their physical separations from the host. Based on this notion we construct a `satellite distribution generator (SDG)' code. Figure \ref{fig:KASmechanism} outlines the SDG. First, we set \italicbold{L} using the position and velocity data of the simulated host--satellite system of interest. Second, we set \italicbold{D} by assigning \textit{d}$_i$ for each orbital pole based on the actual reference system's satellites. 
Last, we generated distributions of satellites sharing identical \italicbold{L} and \italicbold{D} in a circular orbit. 
The orbital phase angles ($\phi$) in a 3-D physical space are known to be quite fragile and become random on a relatively short timescale.
Thus, each satellite is given a random orbital phase angle between 0$^{\circ}$ and 360$^{\circ}$ along the circular track defined by the orbital poles and radial distances. We call the set of modeled geometric systems sharing \italicbold{L} and \italicbold{D} as the `spatially and kinematically analogous system (SKAS)'.

The entire procedure is repeated 10$^5$ times to generate the $c$/$a$ PDF (composed of 10$^5$ SKASs). This procedure is for producing the host--satellite systems sharing the same \italicbold{L} and \italicbold{D}, not reproducing the actual orbit of the satellites. The resulting systems represent the SKASs of the reference host--satellite system of interest. For instance, the SKASs of the MW possess the same \italicbold{L} and \italicbold{D} as observed in the MW. We regard the $c$/$a$ distribution of SKASs of a system as its `intrinsic, underlying $c$/$a$ PDF.' It is important to note that orbital and positional parameters can be intricately intercorrelated because of the astrophysical effects, such as group infall, preferential accretion through large-scale structure like filaments, and triaxial host halo shape. 
Such astrophysical effects produce the present-day \italicbold{L} and \italicbold{D} and are embodied in the SKAS.

\begin{figure*}
\includegraphics[width=18cm]{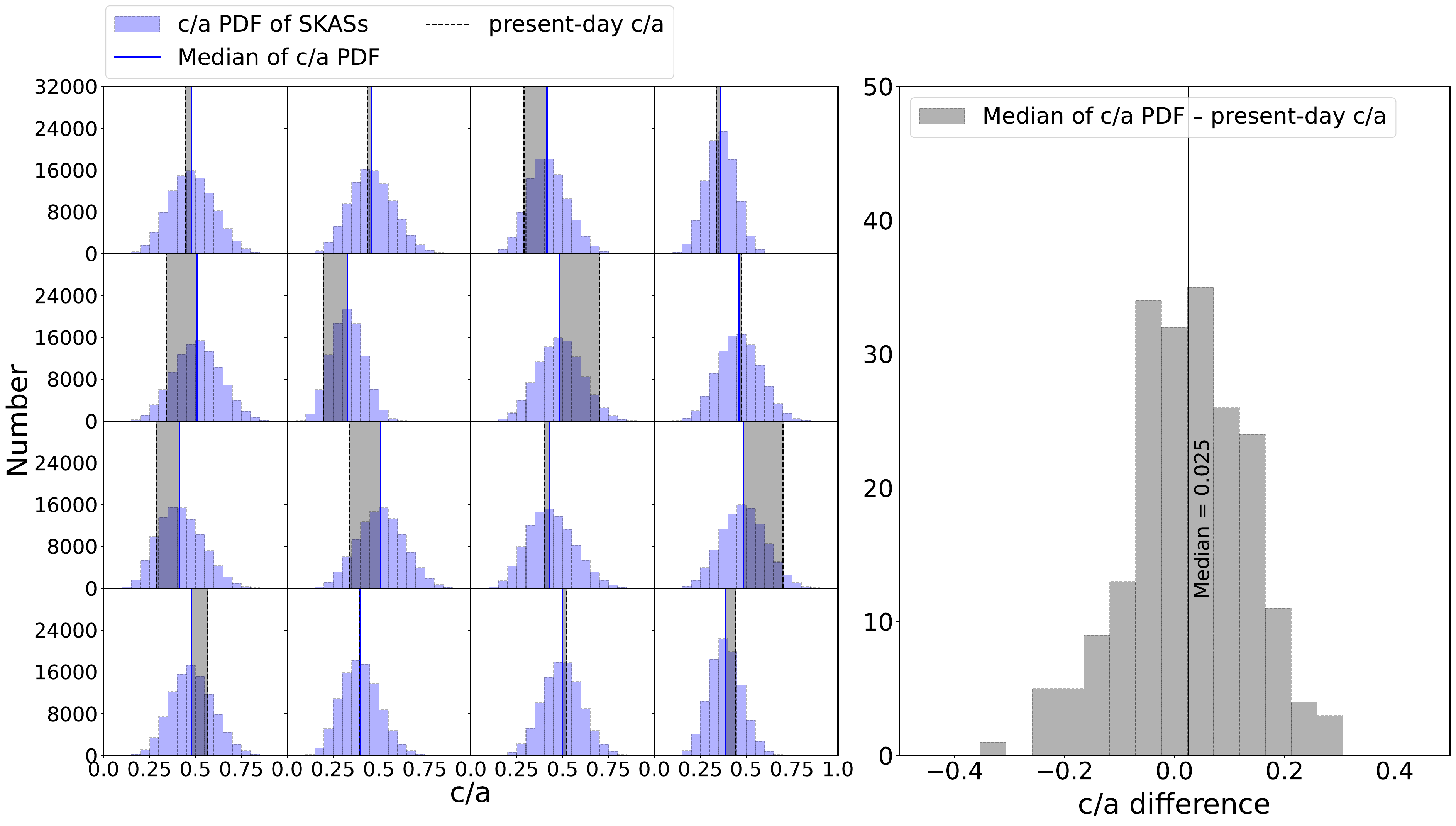}
\centering
\caption{
The difference between the median of the $c$/$a$ PDF and present-day $c$/$a$ for the 10$^5$ SKASs of 202 MW-analogous host--satellite systems in TNG50-1. 
({\it Left}) For the examples of 16 randomly selected systems, we show the intrinsic $c$/$a$ PDFs (blue histograms), their median values (vertical solid blue lines), and the measured $c$/$a$ values (vertical dotted black lines). 
The difference between the median of the $c$/$a$ PDF and measured $c$/$a$ are marked as gray rectangles. 
({\it Right}) We show the distribution of the differences between 
the median of the $c$/$a$ PDF and measured $c$/$a$ for the total 202 MW-analogous host--satellite systems. 
The vertical solid black line at 0.025 indicates the median of the differences.} 
\label{fig:Sanity}
\end{figure*}

Figure \ref{fig:Sanity} presents a sanity check for our SDG and SKAS assumption. If present-day $c$/$a$ values are randomly determined from the $c$/$a$ PDF of the SKASs, the differences between the median of the $c$/$a$ PDFs and simple $c$/$a$ should show a normal distribution centered at zero. We conduct $c$/$a$ calculations for each of the 202 MW-analogous host--satellite systems in the TNG50-1 simulation, obtaining the median values of the $c$/$a$ PDFs. We find that the values of `the median of $c$/$a$ PDF minus the present-day $c$/$a$' for the 202 systems are nearly normally distributed with a median of 0.025. This suggests that present-day $c$/$a$ values are nearly unbiasedly determined from the $c$/$a$ PDF of the SKASs. The marginal discrepancy is attributed to the aforementioned astrophysical effects, which are not fully modeled by the randomized orbital phase angle ($\phi$). We will further discuss this issue in Section \ref{ss:Astrophysicaleffect}.

\subsection{The Rarity of the MW DoS Based on the Intrinsic, Underlying $c$/$a$ PDF}
\label{ss:MWraritywithKAS}

\begin{figure*}
\includegraphics[width=18cm]{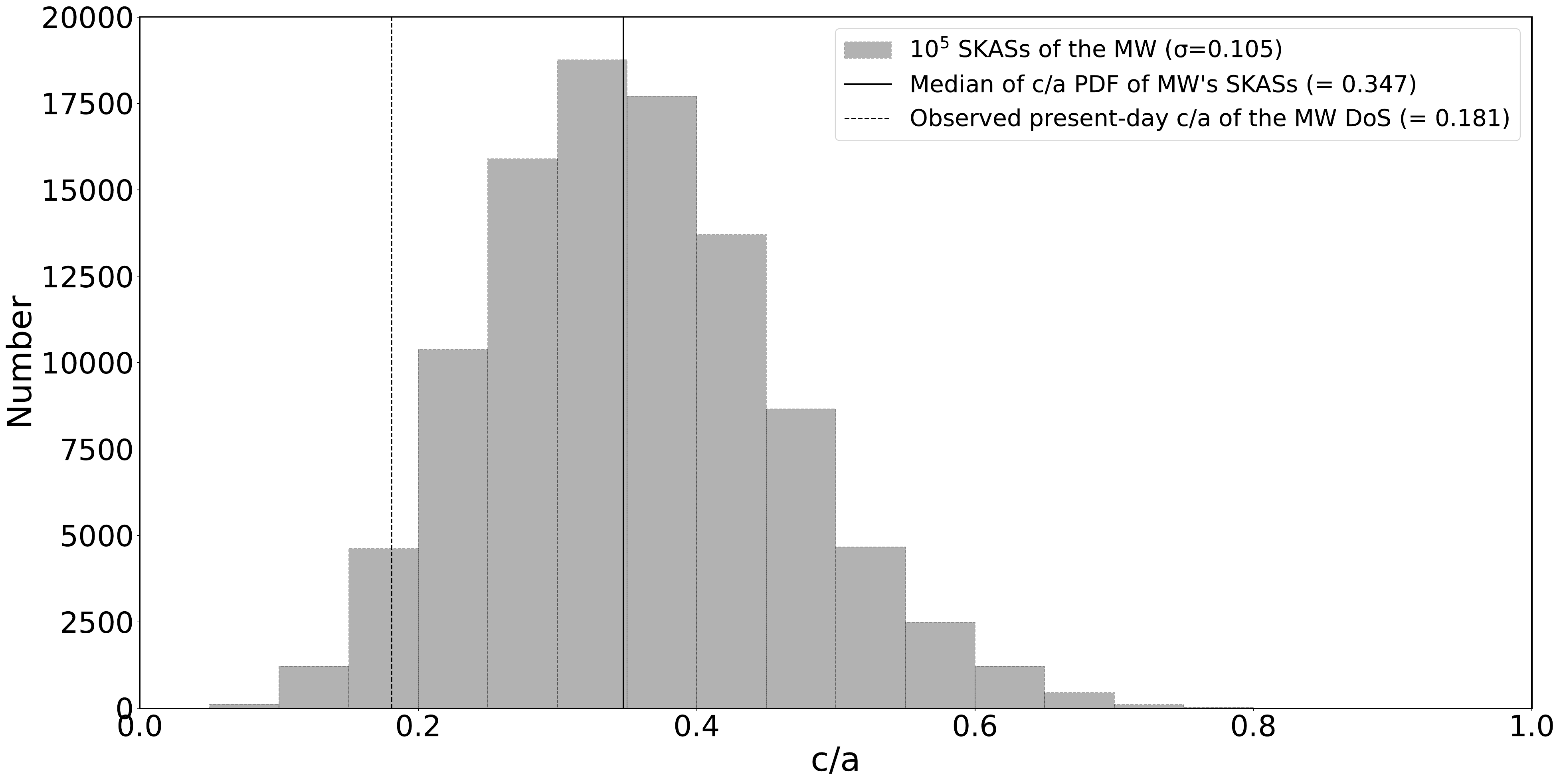}
\centering
\caption{
The $c$/$a$ PDF of 10$^5$ SKASs for the MW DoS. 
The x-axis is for $c$/$a$ and the y-axis is for the number of SKASs within the corresponding $c$/$a$ bins. 
The $c$/$a$ PDF of the MW DoS is quite broad ($\sigma_{c/a}$\,$\sim$\,0.105), implying that a simple present-day $c$/$a$ value, combined with its highly time-variable nature, cannot fully represent the degree of flatness. 
The vertical solid black line indicates the median of the $c$/$a$ PDFs at 0.347. 
The vertical dotted black line represents the observed present-day $c$/$a$ of the MW DoS at 0.181, which is $\sim$1.8\,$\sigma$ way from the median of the $c$/$a$ PDFs.
} 
\label{fig:MWKAS}
\end{figure*}

Figure \ref{fig:MWKAS} shows the $c$/$a$ PDF of the MW DoS, consisting of the 10$^5$ SKASs. We find that the $c$/$a$ PDF of the MW DoS is quite broad ($\sigma_{c/a}$\,$\sim$\,0.105), implying that a simple present-day $c$/$a$ value, combined with its highly time-variable nature \citep[e.g.,][]{2019MNRAS.488.1166S, 2023ApJ...954..128X}, cannot fully represent the degree of flatness. 
The vertical solid black line indicates the median of the $c$/$a$ PDFs at 0.347. The vertical dotted black line represents the observed present-day $c$/$a$ of the MW DoS at 0.181, which is $\sim$1.8\,$\sigma$ way from the median of the $c$/$a$ PDFs.
We further emphasize that relying solely on present-day $c$/$a$ should be avoided because it does not consider the effect of the two flatness-affecting factors: the orbital pole alignment and radial concentration of satellites. 
For instance, a few by-chance outlier satellites far away by chance from the host galaxy can thicken the host--satellite system even though they have the MW-like orbital pole configuration \citep{2019MNRAS.488.1166S}. In a similar vein, a few by-chance outliers having different orbital poles from the rest can thicken the systems even under the MW-like radial configuration.

\begin{figure*}
\includegraphics[width=18cm]{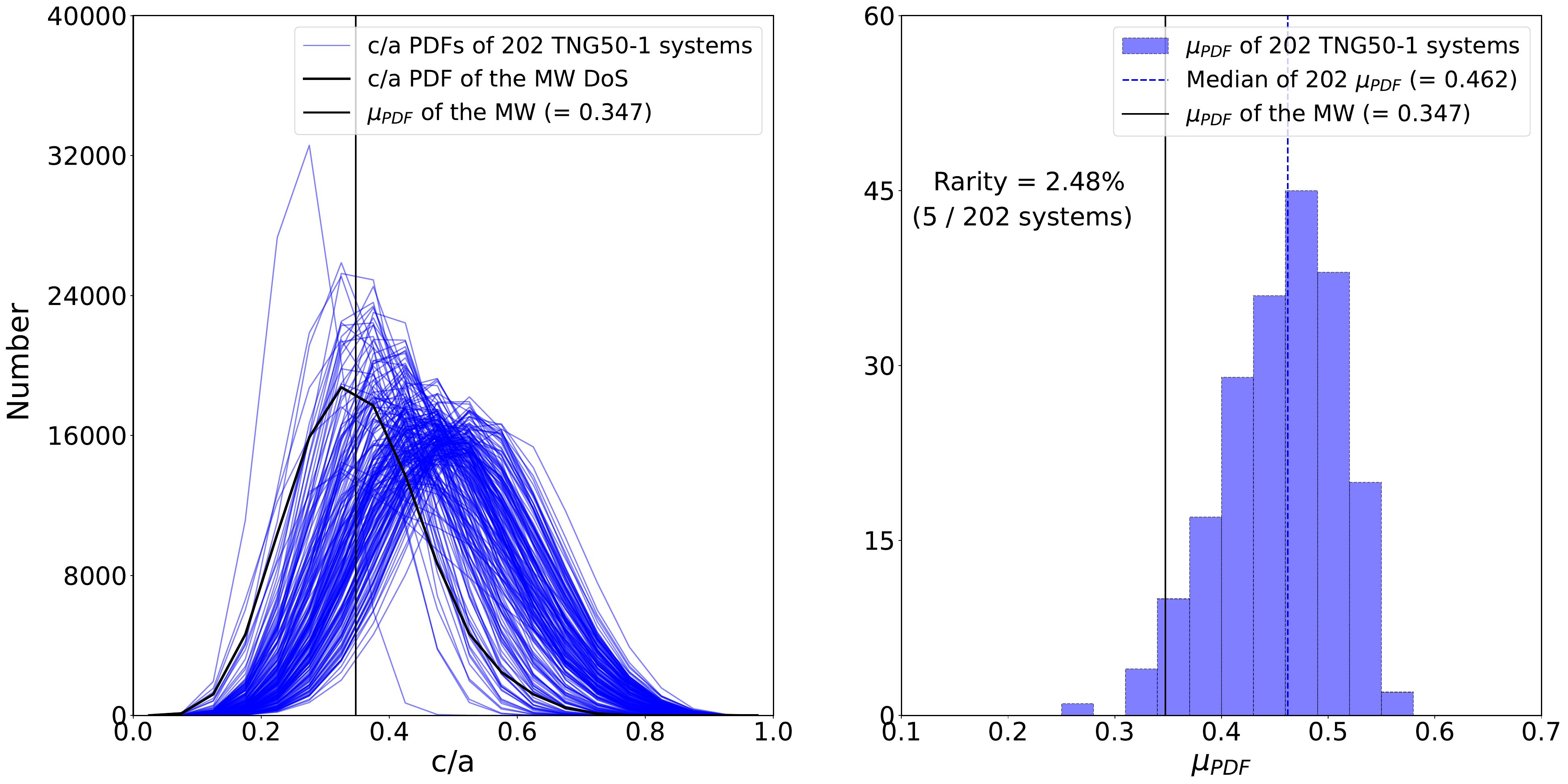}
\centering
\caption{
The $c$/$a$ PDFs ({\it Left}) and their corresponding $\mu_{\rm PDF}$ ({\it Right}) for the MW and 202 MW-analogous host--satellite systems of the TNG50-1 simulation.
({\it Left}) We show the intrinsic $c$/$a$ PDFs of 202 systems in TNG50-1 (thin blue loci) and the MW DoS (thick black locus). 
All $c$/$a$ PDFs are quite broad.
The vertical solid black line is for $\mu_{\rm PDF}$ of the MW DoS at 0.347.
({\it Right}) We show the $\mu_{\rm PDF}$ distribution of all 202 systems. 
The vertical blue dotted line is for the median of $\mu_{\rm PDF}$ of 202 systems at 0.462.
The solid black line is for $\mu_{\rm PDF}$ of the MW DoS at 0.347.
The rarity of the MW DoS based on $\mu_{\rm PDF}$ is calculated to be 2.48\,\% (= 5/202).
} 
\label{fig:intrinsics}
\end{figure*}

To address this limitation, we propose to use the median value of the $c$/$a$ PDF (hereafter, denoted by ``$\mu_{\rm PDF}$'') as a better indicator of a system's flatness. Figure \ref{fig:intrinsics} presents the $c$/$a$ PDFs and $\mu_{\rm PDF}$ for the MW and 202 MW-analogous host--satellite systems of the TNG50-1 simulation. The left panel shows the shape of intrinsic $c$/$a$ PDFs and the right panel shows the corresponding distribution of $\mu_{\rm PDF}$. All $c$/$a$ PDFs are quite broad, implying that the problem of relying solely on present-day $c$/$a$ is ubiquitous. 
The right panel shows the $\mu_{\rm PDF}$ distribution of all 202 systems. The vertical blue dotted line is for the median of $\mu_{\rm PDF}$ of 202 systems at 0.462. The solid black line is for $\mu_{\rm PDF}$ of the MW DoS at 0.347. The rarity of the MW DoS based on $\mu_{\rm PDF}$ is 2.48\,\%.
Overall, the newly calculated rarity of the MW DoS using $\mu_{\rm PDF}$ shows that the MW is still a rare system.

\begin{figure*}
\includegraphics[width=18cm]{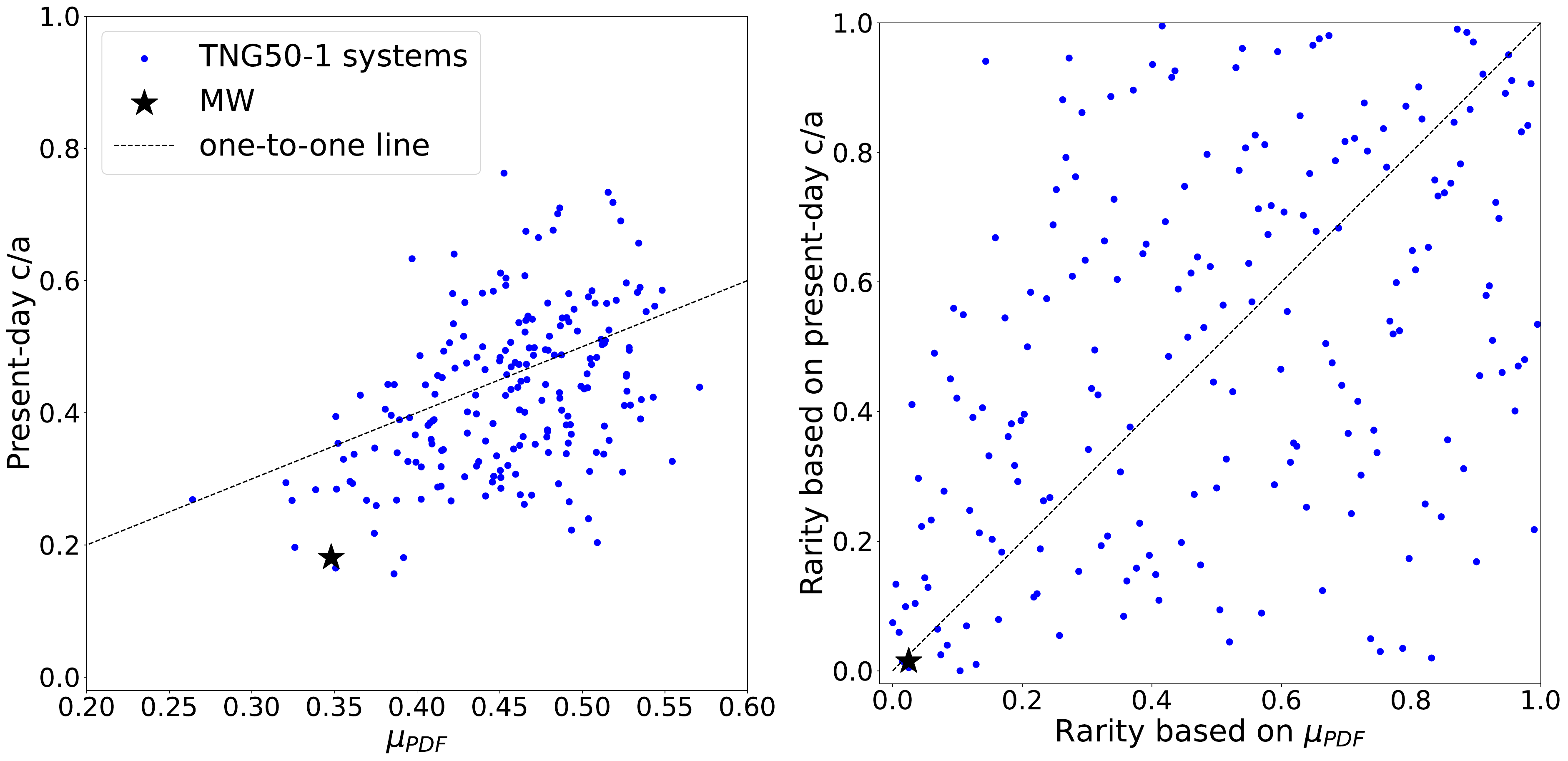}
\caption{
({\it Left}) The relation between $\mu_{\rm PDF}$ and present-day $c$/$a$ of the MW (black star) and the 202 MW-analogous host--satellite systems in TNG50-1 (blue dots).
The dashed black line denotes the one-to-one relation.
Although present-day $c$/$a$ roughly follows $\mu_{\rm PDF}$, the spread is sizeable. 
({\it Right}) 
The relation between the rarity based on $\mu_{\rm PDF}$ (x-axis) and the rarity based on present-day $c$/$a$ (y-axis) for the MW (black star) and the 202 MW-analogous host--satellite systems in TNG50-1 (blue dots). 
The dashed black line denotes the one-to-one relation.
We note that there is no correlation between the two types of rarity. 
The MW (black star) is located at the bottom-left corner of the plot, indicating that it comes out rare in both criteria. 
However, the rarity based on $c$/$a$ should be taken as a by-chance occasion.
}
\label{fig:caandraritytwoplots}
\end{figure*}

The left panel of Figure \ref{fig:caandraritytwoplots} shows the relation between $\mu_{\rm PDF}$ and present-day $c$/$a$ of the MW and the 202 MW-analogous host--satellite systems in TNG50-1. We find that, although present-day $c$/$a$ roughly follows $\mu_{\rm PDF}$, the distribution is very wide. The right panel of Figure \ref{fig:caandraritytwoplots} shows that there is no correlation between the two types of rarity. The MW is located at the bottom-left corner of the plot, indicating that it comes out rare in both criteria. However, the rarity based on $c$/$a$ should be taken as a by-chance occasion.

In short, the present-day $c$/$a$ value is utterly fortuitously selected among the intrinsic, underlying $c$/$a$ PDF. The $c$/$a$ PDF of the MW DoS is quite broad ($\sigma_{c/a}$\,$\sim$\,0.105), implying that a simple present-day $c$/$a$ value, combined with its highly time-variable nature \citep[e.g.,][]{2019MNRAS.488.1166S, 2023ApJ...954..128X}, does not properly represent the degree of flatness. Based on the intrinsic $c$/$a$ PDF, we re-evaluate the rarity of the MW DoS by comparing it with IllustrisTNG and find that even with the new measure of flatness, the MW DoS remains highly uncommon.

\section{What Causes the Extreme Rareness of the MW D\lowercase{o}S?} \label{sec5}

We have demonstrated that the MW DoS is highly uncommon, even when considering the $c$/$a$ PDF. This section aims to parameterize the factors influencing the flatness of a host--satellite system and explore how the parameters shape the $c$/$a$ PDF. This will give insights into the underlying cause of the MW DoS being uncommon in the $\Lambda$CDM cosmology.

\subsection{Parameterization of the Orbital Pole Alignment and Radial Concentration}
\label{ss:Parametersforplane-makingfactor}

\begin{figure*}
\includegraphics[width=18cm]{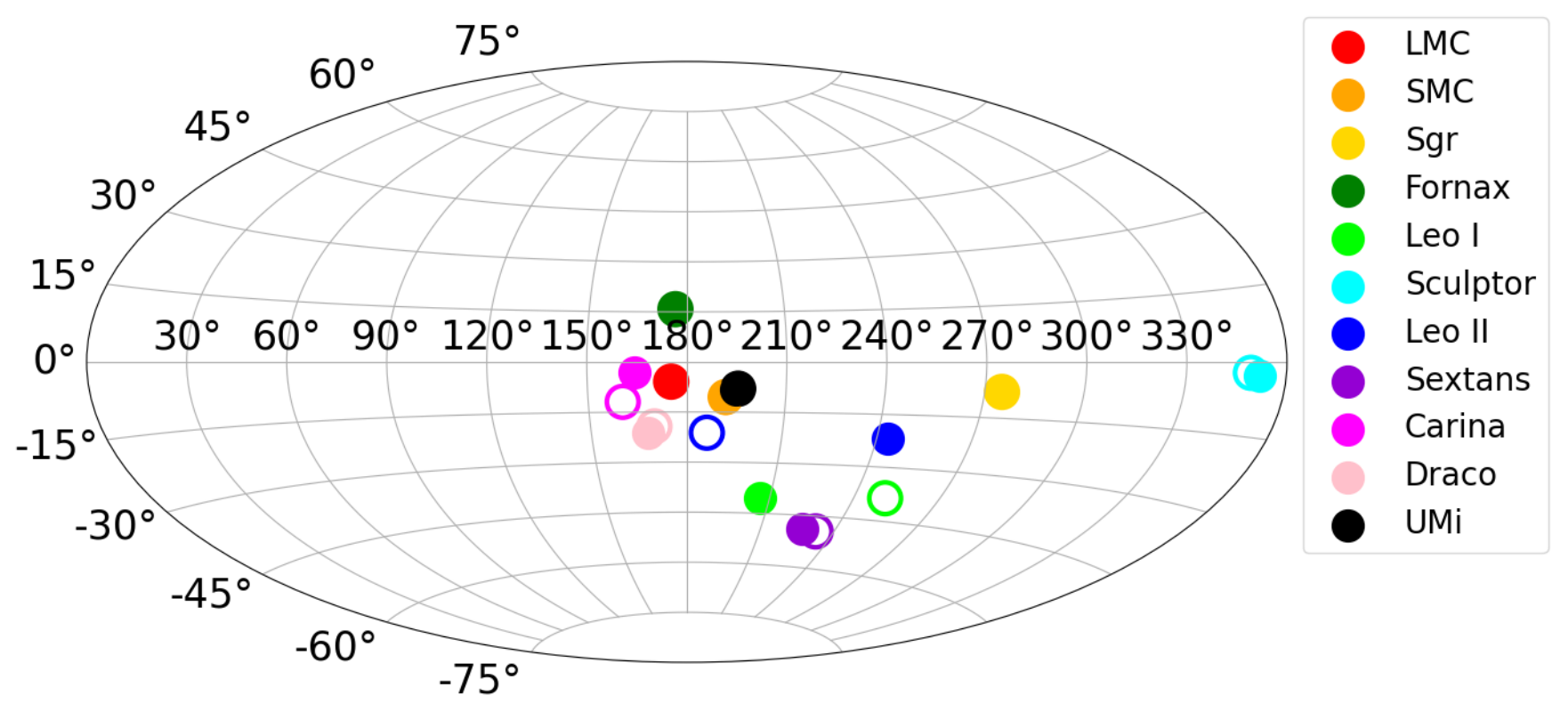}
\centering
\caption{Aitoff projection plot of the orbital poles of 11 classical satellites of the MW. Each circle represents the galactic latitude ($b$) and longitude ($l$) of a satellite's orbital pole. Hollow circles indicate the orbital poles from \citet{2019MNRAS.488.1166S}, while filled circles represent the orbital poles used in this study, sourced from \citet{2020MNRAS.491.3042P}, \citet{2024A&A...681A..73T}, and \citet{2024ApJ...971...98B}. We note that $min$($\alpha_{8}$) increases from 22° to 31.6° due to the updated velocity data affecting the orbital poles of Leo I and Leo II.} 
\label{fig:twoaitoff}
\end{figure*}

To quantify the degree of orbital pole alignment, we adopt the approach introduced by \citet{2019MNRAS.488.1166S}, employing the minimum opening angle, $min$($\alpha_{N}$), that encompasses $N$ out of the 11 orbital poles of the satellites. A smaller value of $min$($\alpha_{N}$) signifies higher kinematic coherence and better alignment of the orbits. The $\alpha_{N}$ values are calculated using 12,288 equally spaced HEALPIX{\footnote{http://healpix.sourceforge.net}} points \citep{2005ApJ...622..759G, Zonca2019}. For the MW satellite system, the minimum opening angle, $min$($\alpha_{8}$), involving eight out of the 11 orbital poles is employed as a measure of kinematic coherence. The MW's kinematic coherence, $min$($\alpha_{8}$), is determined to be 31.6°, approximately 10° larger than the previous value of 22° reported by \citet{2019MNRAS.488.1166S}. This discrepancy arises from the difference in the orbital pole data, particularly for Leo I and Leo II. Figure \ref{fig:twoaitoff} presents the orbital pole distribution from \citet{2019MNRAS.488.1166S} and the updated orbital poles used in this study. The increase in $min$($\alpha_{8}$) occurs as Leo II moves away from the orbital pole cluster, while Leo I moves closer to the cluster. The updated data narrows the opening angle gap between the 8th and 9th satellites from 40° to 16°. But still, the opening angle first increases sharply from the 8th to 9th satellites. Hence, we continue to use $min$($\alpha_{8}$) in our analysis.

On the other hand, we parameterize the degree of radial concentration of a satellite system by the normalized mean distance from the host ($\langle d \rangle_{\rm norm}$). The parameter represents the mean of the distances of 11 satellites divided by the farthest satellite's separation, ranging from $\sim$\,0.09 (indicating that, except for one satellite, all others are on top of the host) to 1.0 (indicating that all satellites are equidistant from the host). A smaller value indicates a stronger central concentration. For the MW DoS, $\langle d \rangle_{\rm norm}$ is calculated to be 0.432.

\subsection{Breaking Down the Underlying Causes of the Rareness of the MW DoS}
\label{ss:MWlikeDoSraritycause}

\begin{figure*}
\includegraphics[width=18cm]{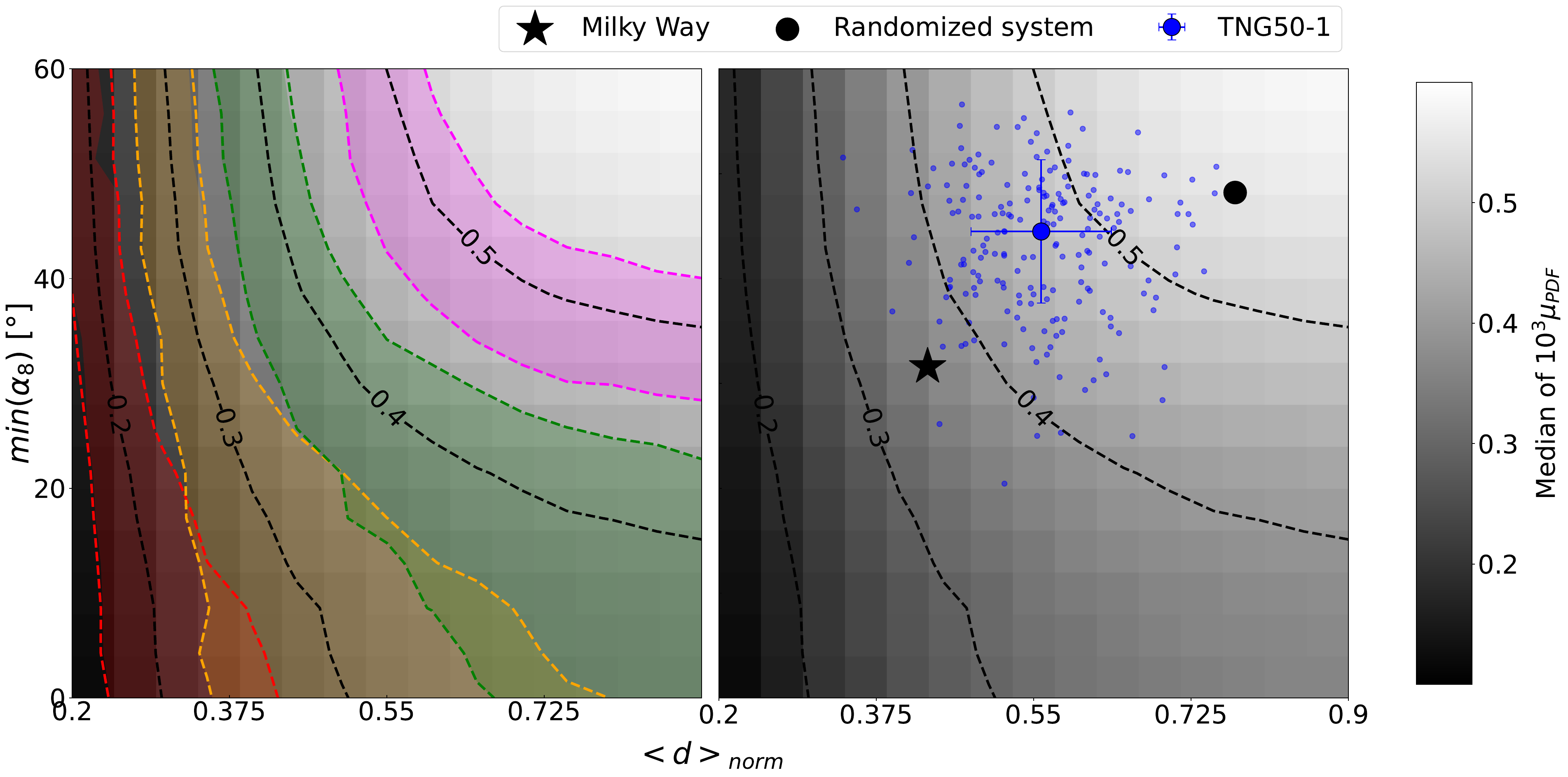}
\centering
\caption{
The $\mu_{\rm PDF}$ behavior of the 202 MW-analogous host--satellite systems in TNG50-1 as a function of the two parameters: $min$($\alpha_{N}$) and $\langle d \rangle_{\rm norm}$. 
The x-axis represents the parameterization of the radial concentration of satellites ($\langle d \rangle_{\rm norm}$), where smaller values indicate a more center-concentrated system. 
The y-axis represents the parameter for orbital pole alignment ($min$($\alpha_{8}$)), with smaller opening angles indicating better-aligned orbital poles (higher kinematic coherence).
We break down the parameters into 15 smaller intervals. 
For each $min$($\alpha_{N}$) versus $\langle d \rangle_{\rm norm}$ grid, we generate $\mu_{\rm PDF}$ of SKASs for 10$^3$ random host--satellite systems and take their median value.
The white-to-black color bar on the right illustrates the median of 10$^{3}$ $\mu_{\rm PDF}$.
({\it Left}) The contours (dotted lines) show the iso-$c$/$a$ loci with the 1$\sigma$ uncertainty bands. 
Note that a more aligned set of orbital poles and a more centrally concentrated satellite distribution contribute to a flatter structure. 
({\it Right}) The same as the left panel but shows our data without uncertainty bands for clarity. 
The locations of the TNG50-1 systems (small blue dots and a large blue circle with an error bar), the MW (black large star), and the full-randomized system  (black large circle) are denoted. 
It is evident that both parameters of the MW DoS significantly favor a planar configuration compared to those of the TNG50-1 host--satellite systems. 
} 
\label{fig:finalplot}
\end{figure*}

Figure \ref{fig:finalplot} shows the $\mu_{\rm PDF}$ behavior of the 202 MW-analogous host--satellite systems in TNG50-1 as a function of the two parameters: $min$($\alpha_{N}$) and $\langle d \rangle_{\rm norm}$. We take 15 evenly distributed grids for each parameter. 
For each $min$($\alpha_{N}$) versus $\langle d \rangle_{\rm norm}$ grid, we generate $\mu_{\rm PDF}$ of SKASs for 10$^3$ random host--satellite systems and take their median value.
In the left panel, the contours (dotted lines) show the iso-$c$/$a$ loci, which demonstrate that a more aligned set of orbital poles and a more centrally concentrated satellite distribution have a flatter DoS structure.
Concerning the 1\,$\sigma$ uncertainties (denoted by the bands), outlier satellites that have significantly different orbital poles from the rest satellites and are positioned at a greater radial distance will cause bias the $c$/$a$ PDF towards a thicker configuration \cite[e.g.,][]{2019MNRAS.488.1166S}. Conversely, if these orbital pole outliers have a smaller radial distance, the $c$/$a$ PDF shifts towards a thinner configuration.

The right panel is the same as the left panel but shows our data without 1\,$\sigma$ uncertainty bands for clarity. 
The locations of the TNG50-1 systems and the MW are denoted, along with the fully randomized imaginary system. The randomized system is obtained based on 11 imaginary satellites with position vectors (selected randomly from a uniform distribution within the volume of a unit sphere) and velocity vectors (selected randomly from a uniform distribution on the surface of the unit sphere). 
On the one hand, the TNG50-1 systems are on average away from the fully randomized system. This is due to the `astrophysical' effect. While the $min$($\alpha_{8}$) value is similar to the randomized system, the $\langle d \rangle_{\rm norm}$ value for the TNG50-1 system is considerably smaller than that of the randomized system. On the other, in the case of the MW DoS, there are substantial differences from the fully randomized system in both the orbital pole alignment and radial concentration. It is also evident that, for the MW DoS, both parameters serve to deviate considerably from the TNG50-1 systems. As a result, the MW DoS is 2.49\,$\sigma$ away from the median of the TNG50-1 systems in the two-parameter space, which gives the rarity of the MW DoS to be 0.64\,\%.

In short, we confirm that {\it both} the positional coherence (i.e., the central concentration represented by $\langle d \rangle_{\rm norm}$) and kinematic coherence (i.e., the orbital pole alignment by $min$($\alpha_{8}$)) contribute simultaneously to the rarity of the MW DoS. The findings strongly suggest that both parameters influencing the flatness of a host--satellite system must conspire accurately to make the MW DoS. 

\section{Discussion} \label{sec6}

\subsection{Effect of the Different Choices of the Satellite Selection Criteria}
\label{ss:sampleselect}

As explained in Section \ref{ss:Simulation}, the satellites are selected according to three criteria, one of which is that they should be bound to the host or sufficiently strongly interacting to eventually merge. 
This is determined by calculating the total energy of the host-satellite two-body system.
In this section, we evaluate how much change the total energy criterion brings. 
To this, we exploit {\it all} SUBFIND halos within the same FoF group of a given host.
In addition, we recall that the radial distance of satellites affects $c$/$a$ (Section \ref{sec5}). 
Thus, we also test the case of all SUBFIND halos that are cut within 1.3$R_{\rm vir}$ to match the observation. 
The observed volume of the MW’s classical 11-satellite system is estimated using the distance of 258 kpc to the farthest classical satellite, Leo I, and the MW's virial radius of $\sim$\,200 kpc \cite[e.g.,][]{2006MNRAS.369.1688D, 2017ApJ...836...92D}. In short, we compare three cases: ($a$) our fiducial setting, i.e., all SUBFIND halos with the total energy cut\footnote{For the case ($a$), the satellites take a volume in a way that 90\% are within $1.4 R_{\rm vir}$ and 99\% are within $2.1 R_{\rm vir}$.}, ($b$) all SUBFIND halos within the same FoF group\footnote{For the case ($b$), the satellites take a volume in a way that 90\% are within $1.5 R_{\rm vir}$ and 99\% are within $2.2 R_{\rm vir}$.}, and ($c$) all SUBFIND halos confined within 1.3$R_{\rm vir}$.

\begin{figure*}
\includegraphics[width=18cm]{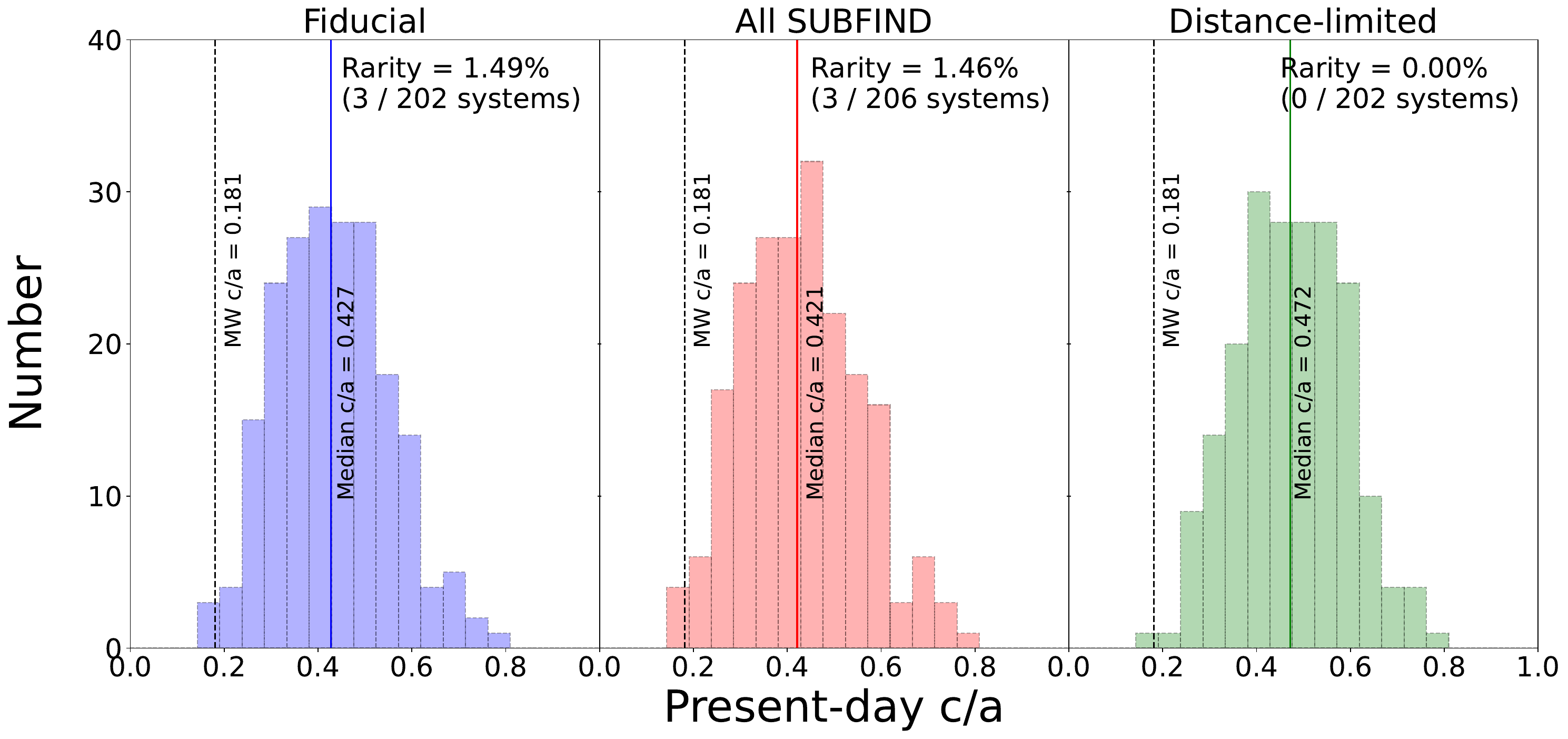}
\centering
\caption{The same as Figure \ref{fig:presentca}, but for two other satellite selection methods. Each panel shows the present-day $c$/$a$ distribution for our fiducial selection method (left), all SUBFIND selection method (middle), and the distance-limited selection method (right).} 
\label{fig:presentca_threemethod}
\end{figure*}

\begin{figure*}
\includegraphics[width=18cm]{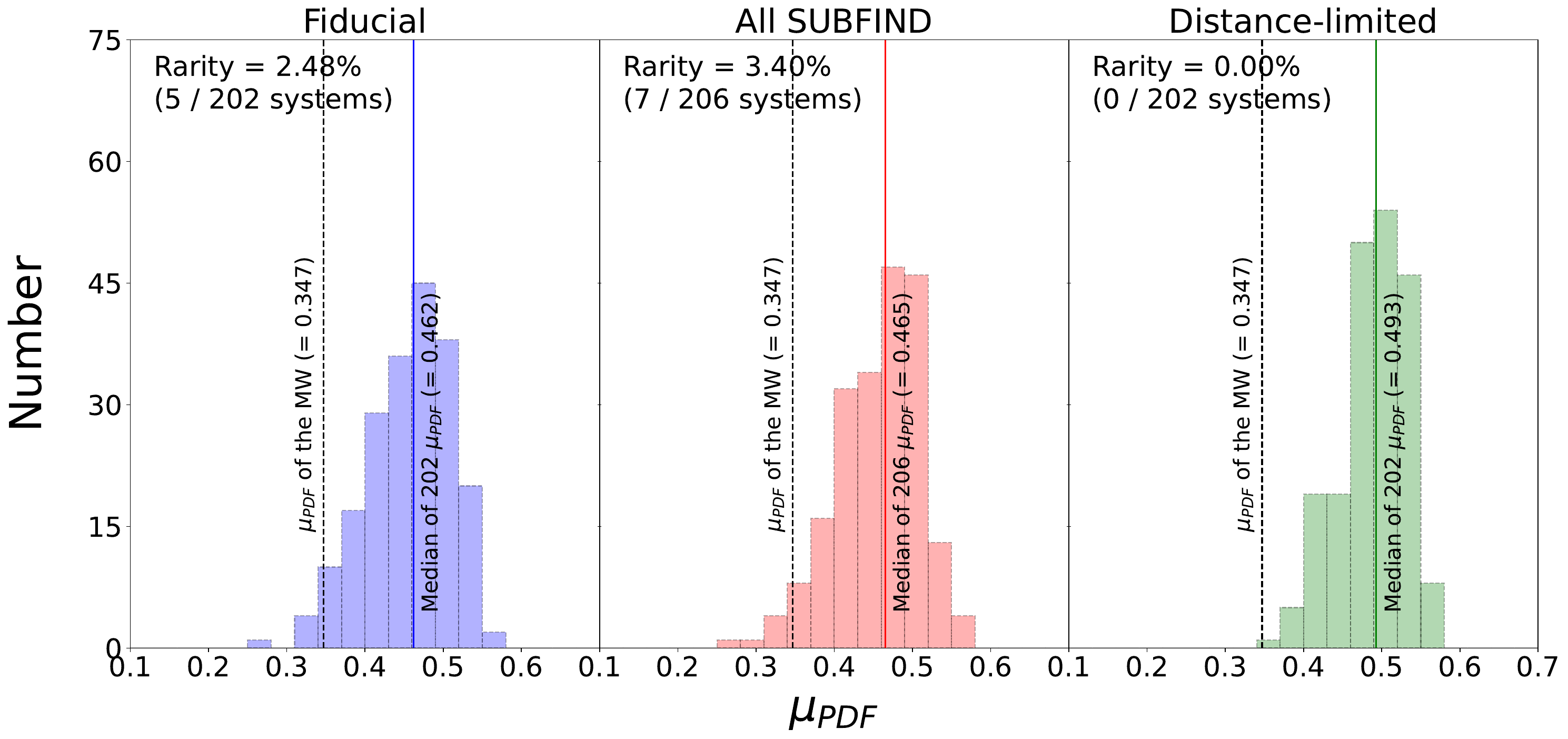}
\centering
\caption{The same as Figure \ref{fig:presentca_threemethod}, but showing the distribution of $\mu_{\rm PDF}$.} 
\label{fig:intrinsicca_threemethod}
\end{figure*}

Figures \ref{fig:presentca_threemethod} and \ref{fig:intrinsicca_threemethod} show the present-day $c$/$a$ and $\mu_{\rm PDF}$ distributions for the MW DoS for three different cases. 
Dropping the total energy criterion (the case ($a$) to case ($b$)) increases the number of MW-analogous host--satellite systems by $\sim$\,2\,\% and the number of subhalos classified as `satellites' by $\sim$\,1\,\%. 
This difference is comparable to the result of \citet{2013MNRAS.436.1765M}, who found that $\sim$\,2\,\% of host--satellite pairs are not bound. 
On the other hand, adding the 1.3$R_{\rm vir}$ cut (the case ($b$) to case ($c$)) leads to a $\sim$\,2\,\% decrease in the number of MW-analogous host--satellite systems.

More importantly, it turns out that the introduction of the total energy criterion (case ($a$), blue histogram) brings about only minor changes to the SKAS of MW-like systems compared to the case of all SUBFIND objects (case ($b$), red histogram). 
There are a slight increase in the present-day $c$/$a$ and a slight decrease in $\mu_{\rm PDF}$. 
While the increase in present-day $c$/$a$ has little impact on the rarity of the MW DoS, the change in $\mu_{\rm PDF}$ leads to a small decrease in the rarity of the MW DoS by 1\%.
On the other hand, adding the 1.3$R_{\rm vir}$ cut to all SUBFIND halos (case ($c$), green histogram) decreases the rarity of the MW DoS even to 0.00\% (0 out of 202 systems) as the median of present-day $c$/$a$ of the SKAS of MW-like systems increases to 0.472. Similarly, for $\mu_{\rm PDF}$, the median increases to 0.493, and the MW DoS rarity decreases even to 0.00\% (0 out of 202 systems).
For comparison, relieving the criterion to the 1.5$R_{\rm vir}$ cut (not shown in the figure) leads to the rarity of the MW DoS to 0.43\% (1 out of 233 systems) as the median of present-day $c$/$a$ becomes 0.465. 
Similarly, for $\mu_{\rm PDF}$, the median is 0.483 and the rarity of the MW DoS is 0.86\% (2 out of 233 systems).

\begin{figure}
\includegraphics[width=9cm]{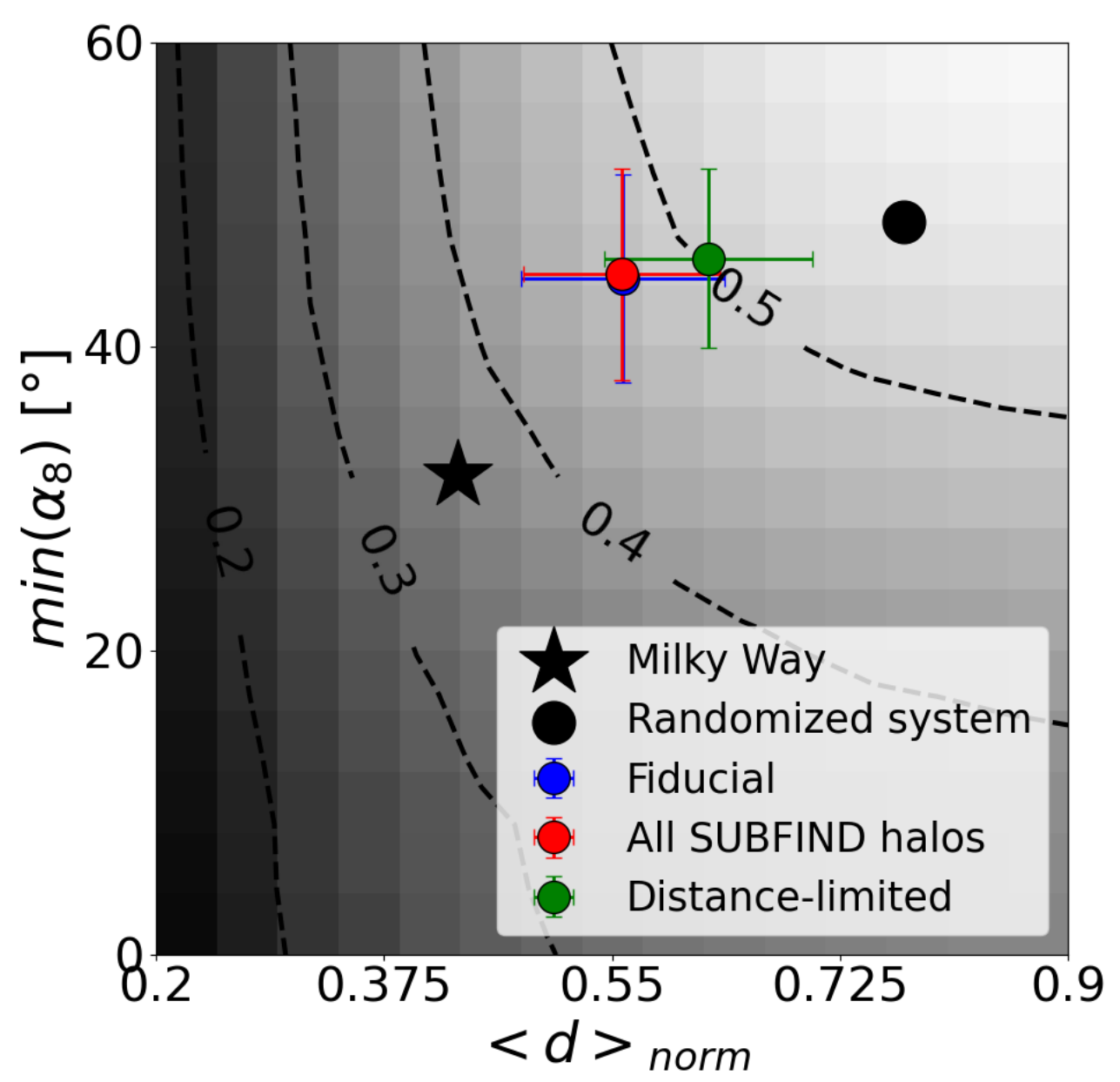}
\centering
\caption{The same as the right panel of Figure \ref{fig:finalplot}, but displaying only the median points and $1\sigma$ dispersion for the three satellite selection methods: Fiducial (blue), all SUBFIND halos (red), and distance-limited (green).} 
\label{fig:twoparam_threemethod}
\end{figure}

Figure \ref{fig:twoparam_threemethod} compares the three different cases in the plane of $\langle d \rangle_{\rm norm}$ versus $min(\alpha_{N})$. 
It turns out that the introduction of the total energy criterion (case ($a$), blue circle) brings about only minor changes to both parameters compared to the case of all SUBFIND halos (case ($b$), red circle). 
For the cases ($a$) and ($b$), the MW DoS is respectively 2.49\,$\sigma$ (see Figure \ref{fig:finalplot}) and 2.52\,$\sigma$ away from the median of the TNG50-1 systems in the two-parameter space, which gives the rarity of the MW DoS to be 0.64\,\% and 0.59\,\%, respectively.
On the other hand, the 1.3$R_{\rm vir}$ cut (case ($c$), green circle) makes the SKASs of MW-like systems have a less center-concentrated distribution (larger $\langle d \rangle_{\rm norm}$), which contributes to the increased $\mu_{\rm PDF}$.
For the cases ($c$), the MW DoS is 3.41\,$\sigma$ away from the median of the TNG50-1 systems in the two-parameter space, which gives the rarity of the MW DoS to be 0.03\,\%.
In summary, when it comes to the $\mu_{\rm PDF}$-based rarity, the case ($c$), compared to the cases ($a$) and ($b$), show higher tension with the observed MW DoS. 
However, the differences do not affect the overall conclusions of this study.

\subsection{Effect of the Heavier Second Most Massive Subhalo}
\label{ss:LMCmasseffect}

As explained in Section \ref{ss:Simulation}, we apply the criterion according to which the second most massive subhalo should not exceed 10\% of the host's total mass, assuming the LMC--MW total mass ratio of 1:10. 
The mass of the LMC, however, remains a topic of ongoing debate.
Some recent studies suggest that the LMC could in fact be considerably more massive, bringing the mass ratio even closer to 1:5 \cite[e.g.,][]{2023Galax..11...59V}. 
In this section, we examine the case for a more massive LMC, i.e., 20\% of the MW total mass.

\begin{figure*}
\includegraphics[width=18cm]{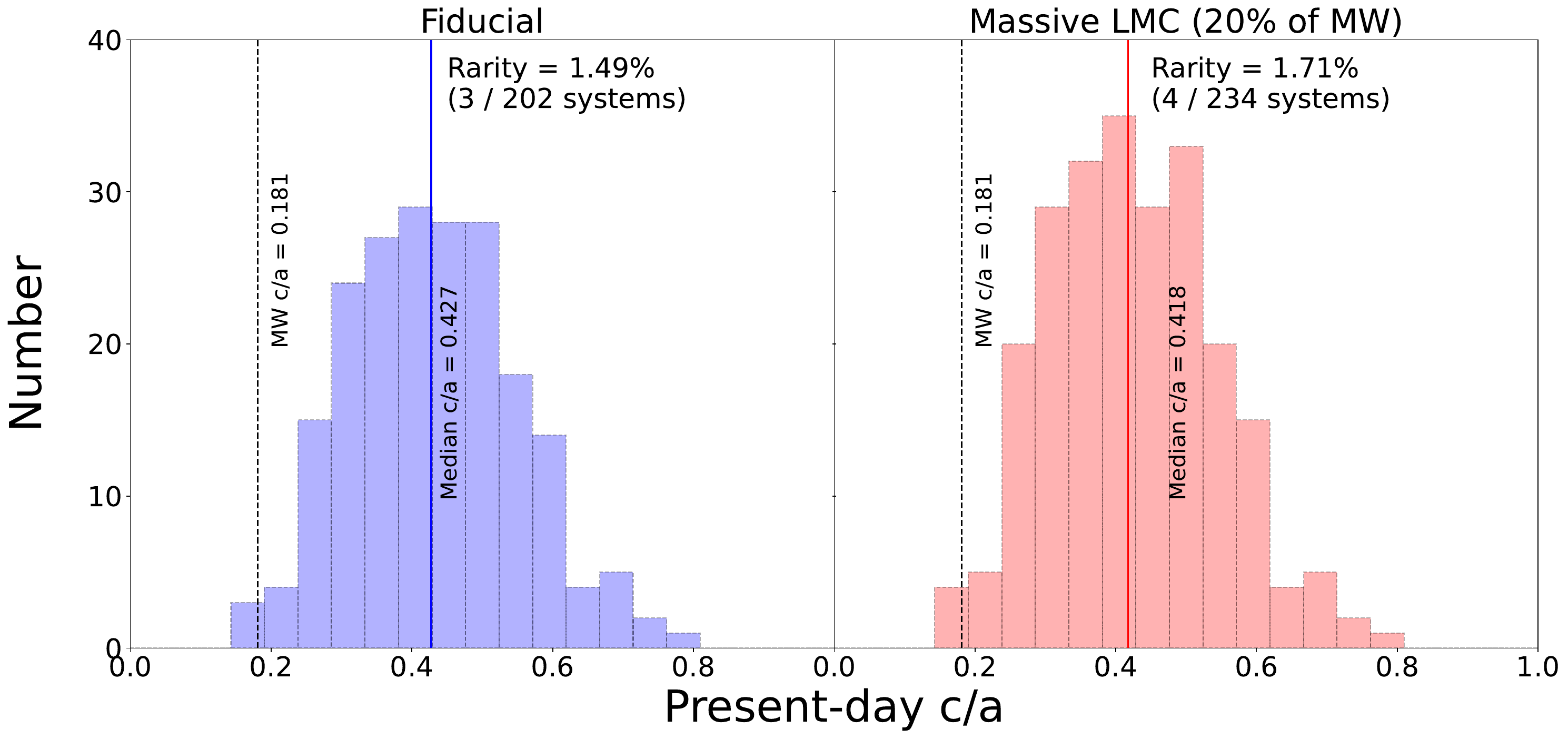}
\centering
\caption{The same as Figure \ref{fig:presentca}, but for increased upper limit for the second most massive halo mass (20\% of the most massive subhalo), reflecting the uncertainty of the LMC's mass estimation. Each panel shows the present-day $c$/$a$ distribution for our fiducial criterion (10\% of the most massive subhalo) (left) and the larger mass criterion (20\% of the most massive subhalo) (right).} 
\label{fig:presentca_largeLMC}
\end{figure*}

\begin{figure*}
\includegraphics[width=18cm]{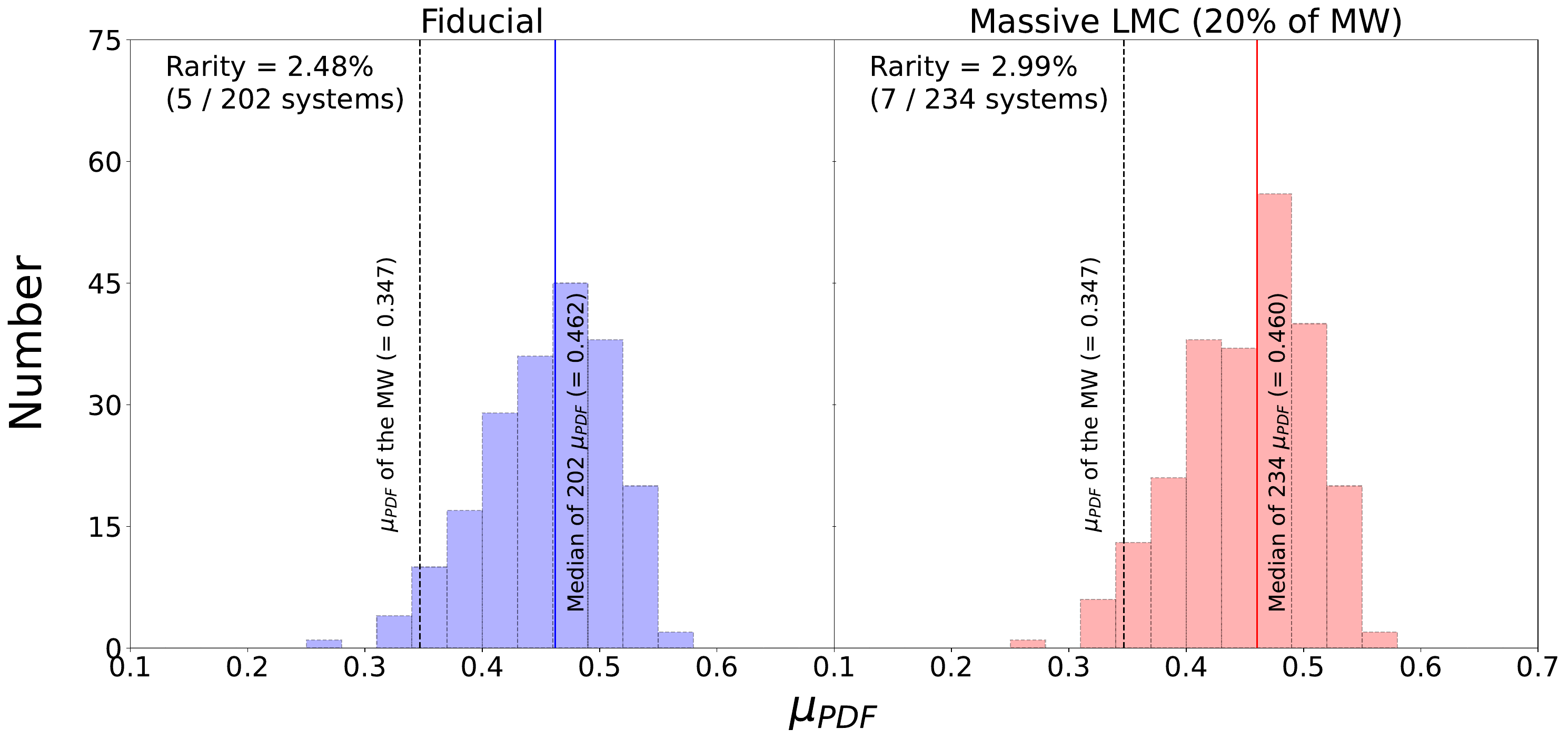}
\centering
\caption{The same as Figure \ref{fig:presentca_largeLMC}, but showing the distribution of $\mu_{\rm PDF}$.} 
\label{fig:intrinsicca_largeLMC}
\end{figure*}

\begin{figure}
\includegraphics[width=9cm]{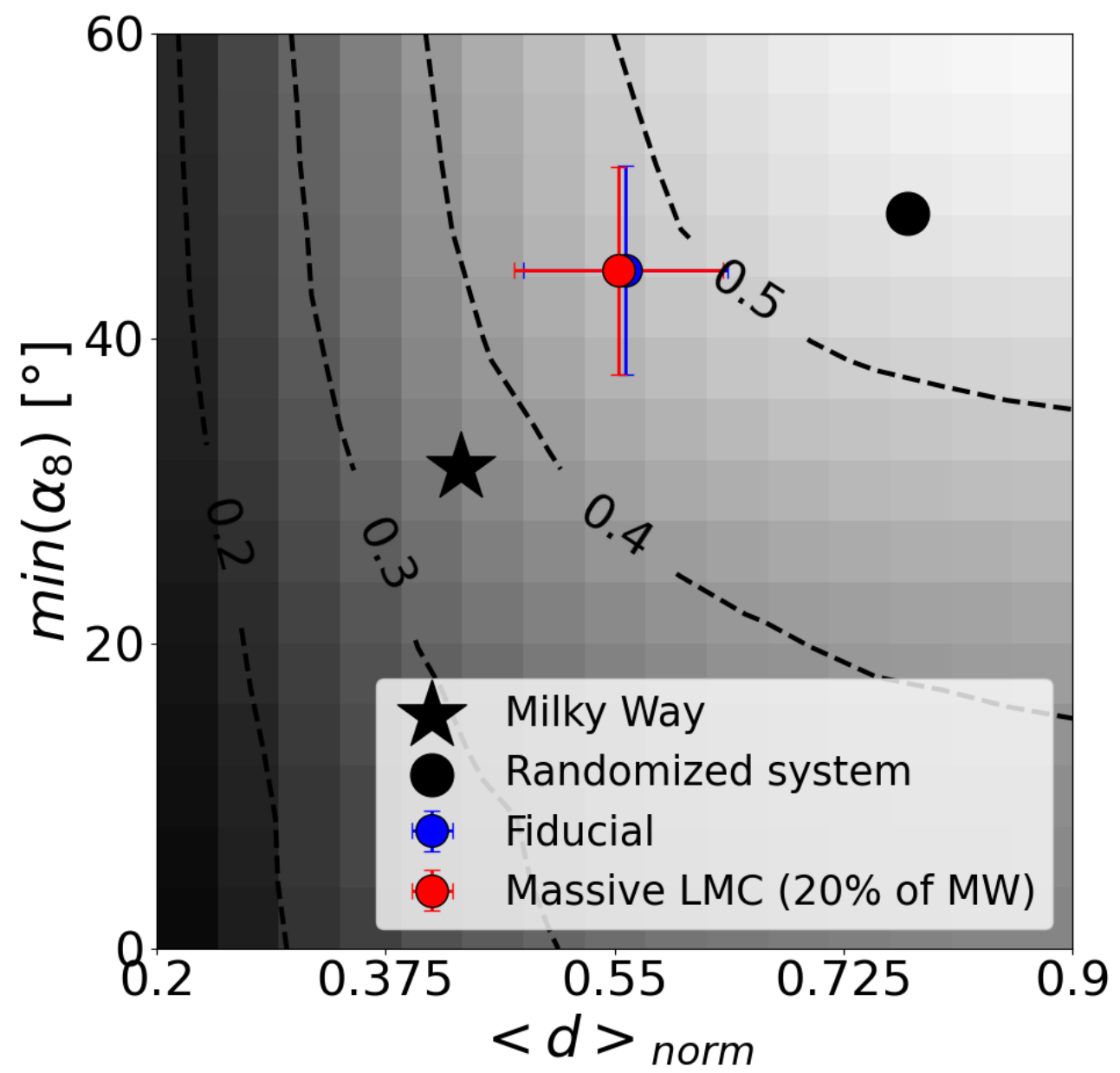}
\centering
\caption{The same as Figure \ref{fig:twoparam_threemethod}, but comparing two different criteria for the second most massive halo: the Fiducial criterion (10\% of the most massive subhalo) in blue, and 20\% of the most massive subhalo in red.} 
\label{fig:twoparam_largerLMC}
\end{figure}

Figures \ref{fig:presentca_largeLMC} and \ref{fig:intrinsicca_largeLMC} show that a heavier second most massive subhalo reduces the median of the present-day $c$/$a$ distribution to 0.418, and the median of $\mu_{\rm PDF}$ remains almost unchanged. 
The rarity of the MW DoS increases to 1.71\% for the present-day $c$/$a$ and to 2.99\% for $\mu_{\rm PDF}$.
Figure \ref{fig:twoparam_largerLMC} shows that the systems with a heavier second most massive subhalo exhibit a slightly smaller $\langle d \rangle_{\rm norm}$, indicating a more center-concentrated distribution. 
This explains the slight increase in the rarity of the MW DoS. 
However, the effect of the heavier second most massive subhalo on kinematic coherence, i.e., $min(\alpha_{8})$, is negligible. 
The MW DoS is 2.43\,$\sigma$ away from the median of the TNG50-1 systems in the two-parameter space, which gives the rarity of the MW DoS to be 0.76\,\%.
In short, a more massive LMC, i.e., 20\% of the MW total mass, does not affect our conclusions.

\subsection{Orbital Plane versus Orbital Direction}
\label{ss:Orbitaldirection}

The methodology proposed in this study does not take into account the fact that the majority of the classical satellite galaxies not only orbit close to a common plane but corotate in a common direction.
Thus, it only considers the orbital plane, not the orbital direction along this plane, which should find less severe tension with simulations \citep[see][]{2020MNRAS.491.3042P}. 
In this section, we discuss the effect of considering the actual orbital directions of the MW satellites in relation to our methodology.

\begin{figure*}
\includegraphics[width=18cm]{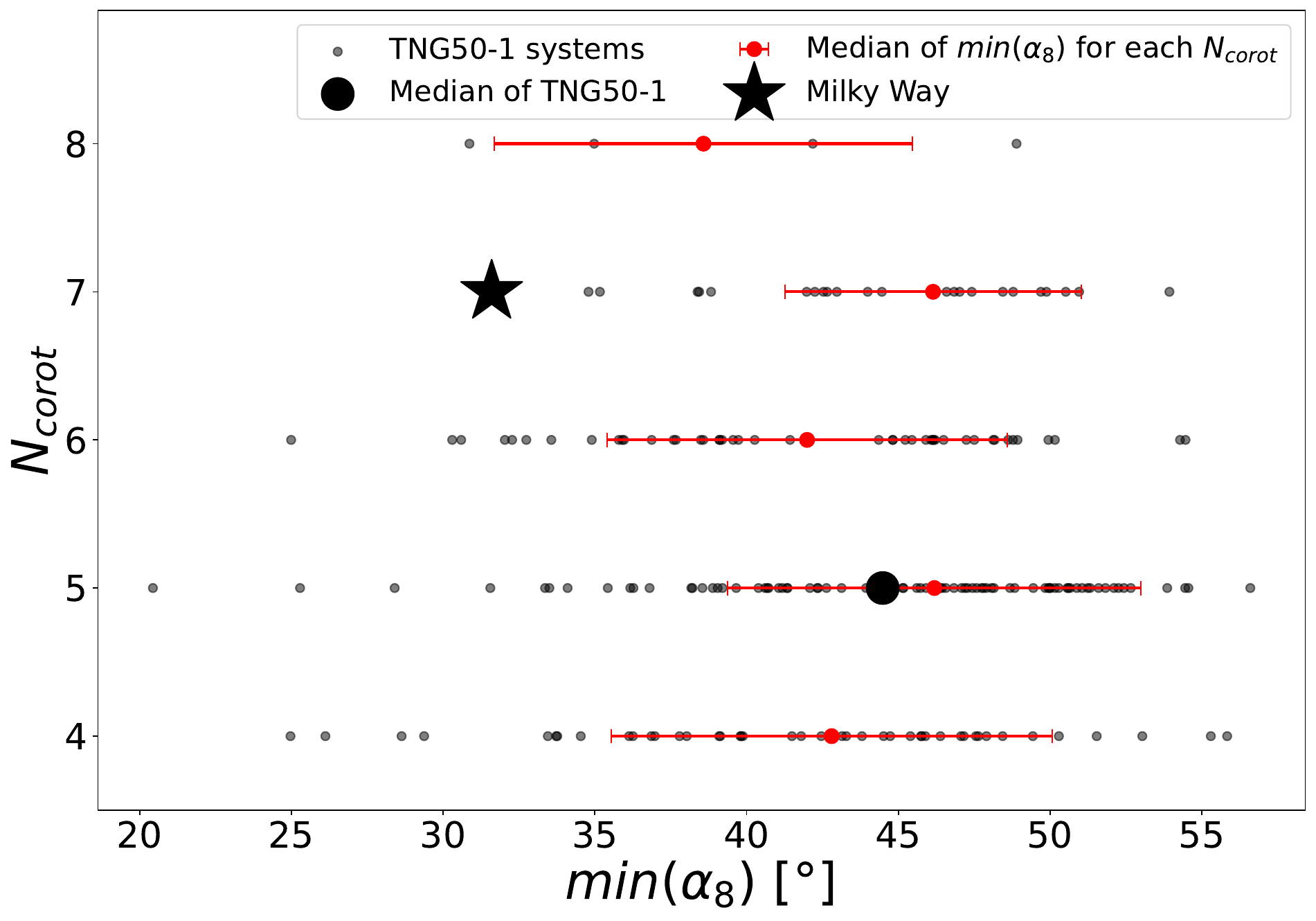}
\centering
\caption{Distributions of kinematic coherence and the number of satellites sharing the same orbital direction. 
The x-axis represents kinematic coherence, parameterized by orbital pole alignment, $min$($\alpha_{8}$). 
The y-axis is for $N_{\rm corot}$, which indicates the number of satellites corotating in the same orbital direction among those whose orbital poles are within $min$($\alpha_{8}$), with larger $N_{\rm corot}$ values implying more satellites moving in the same direction. 
The expected median value for the fully random case is five.
Small grey dots represent TNG50-1 systems, while the large black circle represents the median of these systems. The red error bars at each $N_{\rm corot}$ indicate the median $min$($\alpha_{8}$) and the $1\sigma$ range of TNG50-1 systems for $N_{\rm corot}$ = 4, 5, 6, 7, and 8. The large black star-shaped marker denotes the MW.} 
\label{fig:DirectionVersusPole}
\end{figure*}

In Figure \ref{fig:DirectionVersusPole}, we show $min$($\alpha_{8}$) along with the number of satellites corotating in the same orbital direction ($N_{\rm corot}$) 
among eight satellites (i.e., those whose orbital poles are within $min$($\alpha_{8}$)). 
For an eight-satellite case, the expected median value for a system of a fully random distribution of orbital directions is five.\footnote{ 
For an eight-satellite system of a fully random distribution of orbital directions, the median number of satellites in either direction (drawing with 50\% probability either way) is four. 
However, the median value of $N_{\rm corot}$ should be greater than four because $N_{\rm corot}$ is defined as the larger number of satellites with the same orbital direction. The numbers for the eight-satellite case are 4 (for 4:4), 5 (for 3:5 and 5:3), 6 (for 2:6 and 6:2), 7 (for 1:7 and 7:1), and 8 (for 0:8 and 8:0) and the expected median value for the fully random case is calculated to be five.}
The MW has a notable feature: seven out of eight satellites with concentrated orbital poles (i.e., $< min(\alpha_{8})$) share the same orbital direction. 
By contrast, the median value of $N_{\rm corot}$ for the host--satellite systems in the TNG50-1 turns out to be five, being the same median as a fully random system. 
Remarkably, only one system in the TNG50-1 has both higher $N_{\rm corot}$ and smaller $min$($\alpha_{8}$) than the MW DoS, which gives the rarity of the MW DoS to be 0.50\,\% (= 1/202). 
In short, incorporating the orbital direction increases the tension between the observation and simulation significantly, corroborating \citet{2020MNRAS.491.3042P}.

\subsection{Effect of the Observational Uncertainties on the Rarity of MW DoS}
\label{ss:observationalerror}

As highlighted in \citet{2017AN....338..854P} and \citet{2020MNRAS.491.3042P}, the observational error in satellite velocities can introduce a bias that worsens orbital pole clustering of a given DoS and thus acts to interpret the DoS as being more frequent than it actually is.
For instance, assuming a system of satellites is perfectly flat ($c$/$a$ = 0) with all satellites aligning along a common plane and rotating perfectly within the plane, the $\mu_{\rm PDF}$ distribution would be exactly at zero if there are no errors in satellite velocities. 
With measurement errors, however, the satellites would all show velocities off the perfect plane. 
In turn, their orbital poles would show some scatter that is not intrinsic but due to the errors, leading to shifting the $\mu_{\rm PDF}$ distribution away from zero.
Hence, in light of measurement errors in satellite velocities, the reported rarity of the MW DoS in this study should be taken as an {\it upper limit}, implying that the MW DoS is even more uncommon in conventional cosmology.

Our Monte-Carlo simulation exploiting the MW satellites' velocity errors in the Cartesian coordinate (Table \ref{tab:used data}) shows that the position and shape of the $\mu_{\rm PDF}$ distribution of the MW DoS is highly insensitive to the inclusion of velocity uncertainties, in that the median remains intact (0.347) and the standard deviation changes in the third decimal place (from 0.105 to 0.106). 
However, this simple Monte Carlo sampling of errors in the Cartesian coordinates is inadequate because it does not fully account for the intrinsic internal correlations involved in deriving Cartesian velocities from proper motion data, causing additional artificial dispersion of otherwise more strongly clustered orbital poles.
In upcoming papers, we defer a detailed assessment of how the observational uncertainty of satellite velocities in the {\it proper motion} data affects the rarity estimation of the MW DoS.

\subsection{Consideration of Astrophysical Effects: Towards a better SKAS}
\label{ss:Astrophysicaleffect}

In Section \ref{ss:KAS}, we propose that the SKAS captures present-day characteristics of \italicbold{L} and \italicbold{D}, which are the outcomes of various astrophysical effects.
Thus, the preceding effects should be fairly embodied in the SKAS. 
However, simple modeling \italicbold{L} and \italicbold{D} does not convey the full complexity of reality. 
Moreover, the assumption of random $\phi$ is an important limitation of the SKAS in preserving correlations in satellites' positions and in their orbits.
For instance, several studies showed the anisotropic distribution of satellites (`lopsidedness') around their host galaxies, which may originate from a recently accreted group of satellites \cite[e.g.,][]{2017ApJ...850..132P, 2024MNRAS.529.1405L}.
Also, the limitation results in the bias as shown in Figure \ref{fig:Sanity}.

In our forthcoming papers we will make efforts to elaborate the SKAS scheme. In particular, we will trace simulated halos over time and attempt to take into account the evolution of the SKAS parameters, especially $\phi$ as well as \italicbold{L} and \italicbold{D}.
This will help to better incorporate astrophysical effects, and thus better obtain the intrinsic, underlying $c$/$a$ PDF and the genuine rarity of the MW DoS.

\section{Summary and Conclusion} \label{sec7}

The 11 classical satellite galaxies around the MW show a flattened, disk-like distribution (MW DoS), which is one of the small-scale problems of the $\Lambda$CDM cosmology. Traditionally, the degree of flatness has been gauged using metrics like the minor-major axis ratio ($c$/$a$) or the plane thickness. However, this study questions the validity of the conventional measurement methods and proposes an alternative approach to assess flatness.

The distribution of satellites around a galaxy is mainly dictated by their angular momentum directions and physical separations from the host. We construct the `satellite distribution generator (SDG)' code and generate `spatially and kinematically analogous systems (SKAS)' sharing identical orbital poles and radial distribution. The SKASs disclose the intrinsic, underlying $c$/$a$ PDFs, from which a present-day $c$/$a$ value is utterly fortuitously determined. The SKAS analysis shows that the $c$/$a$ PDF of the MW DoS is quite broad ($\sigma$ $\sim$ 0.105). This, combined with the highly time-variable nature of the present-day $c$/$a$ value, implies that the popular parameter, $c$/$a$, cannot fully represent the intrinsic flatness of a given DoS.

More importantly, the rarity of the MW DoS is re-assessed by comparing it with cosmological simulations (IllustrisTNG50-1). We find that, even with the new measure (i.e, the $c$/$a$ PDF), the MW DoS is highly uncommon in that the rarity ranges 0.00$\sim$3.40\,\% depending on the halo selection method and the fiducial case gives 2.48\,\%. 
We also find that MW satellites’ orbital poles and their separations from the host are simultaneously far more `{\it plane-friendly}' than those of simulated host--satellite systems in IllustrisTNG50-1. Our new approach based on the intrinsic $c$/$a$ PDF corroborates that the MW system is the exception rather than the rule in the $\Lambda$CDM Cosmology, challenging the current structure and galaxy formation model.
The summary of our estimated rarities of the MW DoS using different methodologies is provided in Table \ref{tab:raritysummary}.

\acknowledgments
S.-J.Y. acknowledges support from the Mid-career Researcher Program (No. RS-2024-00344283) and Basic Science Research Program (No. 2022R1A6A1A03053472) through the National Research Foundation (NRF) of Korea. Some of the results in this paper have been derived using the healpy and HEALPix packages.

\bibliography{ref}

\begin{thebibliography}{}
\expandafter\ifx\csname natexlab\endcsname\relax\def\natexlab#1{#1}\fi
\providecommand{\url}[1]{\href{#1}{#1}}
\providecommand{\dodoi}[1]{doi:~\href{http://doi.org/#1}{\nolinkurl{#1}}}
\providecommand{\doeprint}[1]{\href{http://ascl.net/#1}{\nolinkurl{http://ascl.net/#1}}}
\providecommand{\doarXiv}[1]{\href{https://arxiv.org/abs/#1}{\nolinkurl{https://arxiv.org/abs/#1}}}

\bibitem[{{Ahmed} {et~al.}(2017){Ahmed}, {Brooks}, \& {Christensen}}]{2017MNRAS.466.3119A}
{Ahmed}, S.~H., {Brooks}, A.~M., \& {Christensen}, C.~R. 2017, \mnras, 466, 3119, \dodoi{10.1093/mnras/stw3271}

\bibitem[{{An} {et~al.}(2019){An}, {Kim}, {Moon}, \& {Yoon}}]{2019ApJ...887...59A}
{An}, S.-H., {Kim}, J., {Moon}, J.-S., \& {Yoon}, S.-J. 2019, \apj, 887, 59, \dodoi{10.3847/1538-4357/ab535f}

\bibitem[{{Battaglia} {et~al.}(2022){Battaglia}, {Taibi}, {Thomas}, \& {Fritz}}]{2022A&A...657A..54B}
{Battaglia}, G., {Taibi}, S., {Thomas}, G.~F., \& {Fritz}, T.~K. 2022, \aap, 657, A54, \dodoi{10.1051/0004-6361/202141528}

\bibitem[{{Bellazzini} {et~al.}(2013){Bellazzini}, {Oosterloo}, {Fraternali}, \& {Beccari}}]{2013A&A...559L..11B}
{Bellazzini}, M., {Oosterloo}, T., {Fraternali}, F., \& {Beccari}, G. 2013, \aap, 559, L11, \dodoi{10.1051/0004-6361/201322744}

\bibitem[{{Bennet} {et~al.}(2024){Bennet}, {Patel}, {Sohn}, {del Pino Molina}, {van der Marel}, {Libralato}, {Watkins}, {Aparicio}, {Besla}, {Gallart}, {Fardal}, {Monelli}, {Sacchi}, {Tollerud}, \& {Weisz}}]{2024ApJ...971...98B}
{Bennet}, P., {Patel}, E., {Sohn}, S.~T., {et~al.} 2024, \apj, 971, 98, \dodoi{10.3847/1538-4357/ad5349}

\bibitem[{{Buck} {et~al.}(2016){Buck}, {Dutton}, \& {Macci{\`o}}}]{2016MNRAS.460.4348B}
{Buck}, T., {Dutton}, A.~A., \& {Macci{\`o}}, A.~V. 2016, \mnras, 460, 4348, \dodoi{10.1093/mnras/stw1232}

\bibitem[{{Callingham} {et~al.}(2019){Callingham}, {Cautun}, {Deason}, {Frenk}, {Wang}, {G{\'o}mez}, {Grand}, {Marinacci}, \& {Pakmor}}]{2019MNRAS.484.5453C}
{Callingham}, T.~M., {Cautun}, M., {Deason}, A.~J., {et~al.} 2019, \mnras, 484, 5453, \dodoi{10.1093/mnras/stz365}

\bibitem[{{Cautun} {et~al.}(2015{\natexlab{a}}){Cautun}, {Bose}, {Frenk}, {Guo}, {Han}, {Hellwing}, {Sawala}, \& {Wang}}]{2015MNRAS.452.3838C}
{Cautun}, M., {Bose}, S., {Frenk}, C.~S., {et~al.} 2015{\natexlab{a}}, \mnras, 452, 3838, \dodoi{10.1093/mnras/stv1557}

\bibitem[{{Cautun} {et~al.}(2015{\natexlab{b}}){Cautun}, {Wang}, {Frenk}, \& {Sawala}}]{2015MNRAS.449.2576C}
{Cautun}, M., {Wang}, W., {Frenk}, C.~S., \& {Sawala}, T. 2015{\natexlab{b}}, \mnras, 449, 2576, \dodoi{10.1093/mnras/stv490}

\bibitem[{{Chiboucas} {et~al.}(2013){Chiboucas}, {Jacobs}, {Tully}, \& {Karachentsev}}]{2013AJ....146..126C}
{Chiboucas}, K., {Jacobs}, B.~A., {Tully}, R.~B., \& {Karachentsev}, I.~D. 2013, \aj, 146, 126, \dodoi{10.1088/0004-6256/146/5/126}

\bibitem[{{Conn} {et~al.}(2013){Conn}, {Lewis}, {Ibata}, {Parker}, {Zucker}, {McConnachie}, {Martin}, {Valls-Gabaud}, {Tanvir}, {Irwin}, {Ferguson}, \& {Chapman}}]{2013ApJ...766..120C}
{Conn}, A.~R., {Lewis}, G.~F., {Ibata}, R.~A., {et~al.} 2013, \apj, 766, 120, \dodoi{10.1088/0004-637X/766/2/120}

\bibitem[{{Deason} {et~al.}(2011){Deason}, {McCarthy}, {Font}, {Evans}, {Frenk}, {Belokurov}, {Libeskind}, {Crain}, \& {Theuns}}]{2011MNRAS.415.2607D}
{Deason}, A.~J., {McCarthy}, I.~G., {Font}, A.~S., {et~al.} 2011, \mnras, 415, 2607, \dodoi{10.1111/j.1365-2966.2011.18884.x}

\bibitem[{{Dehnen} {et~al.}(2006){Dehnen}, {McLaughlin}, \& {Sachania}}]{2006MNRAS.369.1688D}
{Dehnen}, W., {McLaughlin}, D.~E., \& {Sachania}, J. 2006, \mnras, 369, 1688, \dodoi{10.1111/j.1365-2966.2006.10404.x}

\bibitem[{{Dierickx} \& {Loeb}(2017)}]{2017ApJ...836...92D}
{Dierickx}, M. I.~P., \& {Loeb}, A. 2017, \apj, 836, 92, \dodoi{10.3847/1538-4357/836/1/92}

\bibitem[{{D'Onghia} \& {Lake}(2008)}]{2008ApJ...686L..61D}
{D'Onghia}, E., \& {Lake}, G. 2008, \apjl, 686, L61, \dodoi{10.1086/592995}

\bibitem[{{Errani} {et~al.}(2018){Errani}, {Pe{\~n}arrubia}, \& {Walker}}]{2018MNRAS.481.5073E}
{Errani}, R., {Pe{\~n}arrubia}, J., \& {Walker}, M.~G. 2018, \mnras, 481, 5073, \dodoi{10.1093/mnras/sty2505}

\bibitem[{{Forero-Romero} \& {Arias}(2018)}]{2018MNRAS.478.5533F}
{Forero-Romero}, J.~E., \& {Arias}, V. 2018, \mnras, 478, 5533, \dodoi{10.1093/mnras/sty1349}

\bibitem[{{Fritz} {et~al.}(2018){Fritz}, {Battaglia}, {Pawlowski}, {Kallivayalil}, {van der Marel}, {Sohn}, {Brook}, \& {Besla}}]{2018A&A...619A.103F}
{Fritz}, T.~K., {Battaglia}, G., {Pawlowski}, M.~S., {et~al.} 2018, \aap, 619, A103, \dodoi{10.1051/0004-6361/201833343}

\bibitem[{{Gaia Collaboration} {et~al.}(2018){Gaia Collaboration}, {Helmi}, {van Leeuwen}, {McMillan}, {Massari}, {Antoja}, {Robin}, {Lindegren}, {Bastian}, {Arenou}, {Babusiaux}, {Biermann}, {Breddels}, {Hobbs}, {Jordi}, {Pancino}, {Reyl{\'e}}, {Veljanoski}, {Brown}, {Vallenari}, {Prusti}, {de Bruijne}, {Bailer-Jones}, {Evans}, {Eyer}, {Jansen}, {Klioner}, {Lammers}, {Luri}, {Mignard}, {Panem}, {Pourbaix}, {Randich}, {Sartoretti}, {Siddiqui}, {Soubiran}, {Walton}, {Cropper}, {Drimmel}, {Katz}, {Lattanzi}, {Bakker}, {Cacciari}, {Casta{\~n}eda}, {Chaoul}, {Cheek}, {De Angeli}, {Fabricius}, {Guerra}, {Holl}, {Masana}, {Messineo}, {Mowlavi}, {Nienartowicz}, {Panuzzo}, {Portell}, {Riello}, {Seabroke}, {Tanga}, {Th{\'e}venin}, {Gracia-Abril}, {Comoretto}, {Garcia-Reinaldos}, {Teyssier}, {Altmann}, {Andrae}, {Audard}, {Bellas-Velidis}, {Benson}, {Berthier}, {Blomme}, {Burgess}, {Busso}, {Carry}, {Cellino}, {Clementini}, {Clotet}, {Creevey}, {Davidson}, {De Ridder}, {Delchambre}, {Dell'Oro}, {Ducourant},
  {Fern{\'a}ndez-Hern{\'a}ndez}, {Fouesneau}, {Fr{\'e}mat}, {Galluccio}, {Garc{\'\i}a-Torres}, {Gonz{\'a}lez-N{\'u}{\~n}ez}, {Gonz{\'a}lez-Vidal}, {Gosset}, {Guy}, {Halbwachs}, {Hambly}, {Harrison}, {Hern{\'a}ndez}, {Hestroffer}, {Hodgkin}, {Hutton}, {Jasniewicz}, {Jean-Antoine-Piccolo}, {Jordan}, {Korn}, {Krone-Martins}, {Lanzafame}, {Lebzelter}, {L{\"o}ffler}, {Manteiga}, {Marrese}, {Mart{\'\i}n-Fleitas}, {Moitinho}, {Mora}, {Muinonen}, {Osinde}, {Pauwels}, {Petit}, {Recio-Blanco}, {Richards}, {Rimoldini}, {Sarro}, {Siopis}, {Smith}, {Sozzetti}, {S{\"u}veges}, {Torra}, {van Reeven}, {Abbas}, {Abreu Aramburu}, {Accart}, {Aerts}, {Altavilla}, {{\'A}lvarez}, {Alvarez}, {Alves}, {Anderson}, {Andrei}, {Anglada Varela}, {Antiche}, {Arcay}, {Astraatmadja}, {Bach}, {Baker}, {Balaguer-N{\'u}{\~n}ez}, {Balm}, {Barache}, {Barata}, {Barbato}, {Barblan}, {Barklem}, {Barrado}, {Barros}, {Barstow}, {Bartholom{\'e} Mu{\~n}oz}, {Bassilana}, {Becciani}, {Bellazzini}, {Berihuete}, {Bertone}, {Bianchi}, {Bienaym{\'e}},
  {Blanco-Cuaresma}, {Boch}, {Boeche}, {Bombrun}, {Borrachero}, {Bossini}, {Bouquillon}, {Bourda}, {Bragaglia}, {Bramante}, {Bressan}, {Brouillet}, {Br{\"u}semeister}, {Brugaletta}, {Bucciarelli}, {Burlacu}, {Busonero}, {Butkevich}, {Buzzi}, {Caffau}, {Cancelliere}, {Cannizzaro}, {Cantat-Gaudin}, {Carballo}, {Carlucci}, {Carrasco}, {Casamiquela}, {Castellani}, {Castro-Ginard}, {Charlot}, {Chemin}, {Chiavassa}, {Cocozza}, {Costigan}, {Cowell}, {Crifo}, {Crosta}, {Crowley}, {Cuypers}, {Dafonte}, {Damerdji}, {Dapergolas}, {David}, {David}, {de Laverny}, {De Luise}, {De March}, {de Martino}, {de Souza}, {de Torres}, {Debosscher}, {del Pozo}, {Delbo}, {Delgado}, {Delgado}, {Di Matteo}, {Diakite}, {Diener}, {Distefano}, {Dolding}, {Drazinos}, {Dur{\'a}n}, {Edvardsson}, {Enke}, {Eriksson}, {Esquej}, {Eynard Bontemps}, {Fabre}, {Fabrizio}, {Faigler}, {Falc{\~a}o}, {Farr{\`a}s Casas}, {Federici}, {Fedorets}, {Fernique}, {Figueras}, {Filippi}, {Findeisen}, {Fonti}, {Fraile}, {Fraser}, {Fr{\'e}zouls}, {Gai}, {Galleti},
  {Garabato}, {Garc{\'\i}a-Sedano}, {Garofalo}, {Garralda}, {Gavel}, {Gavras}, {Gerssen}, {Geyer}, {Giacobbe}, {Gilmore}, {Girona}, {Giuffrida}, {Glass}, {Gomes}, {Granvik}, {Gueguen}, {Guerrier}, {Guiraud}, {Guti{\'e}rrez-S{\'a}nchez}, {Hofmann}, {Holland}, {Huckle}, {Hypki}, {Icardi}, {Jan{\ss}en}, {Jevardat de Fombelle}, {Jonker}, {Juh{\'a}sz}, {Julbe}, {Karampelas}, {Kewley}, {Klar}, {Kochoska}, {Kohley}, {Kolenberg}, {Kontizas}, {Kontizas}, {Koposov}, {Kordopatis}, {Kostrzewa-Rutkowska}, {Koubsky}, {Lambert}, {Lanza}, {Lasne}, {Lavigne}, {Le Fustec}, {Le Poncin-Lafitte}, {Lebreton}, {Leccia}, {Leclerc}, {Lecoeur-Taibi}, {Lenhardt}, {Leroux}, {Liao}, {Licata}, {Lindstr{\o}m}, {Lister}, {Livanou}, {Lobel}, {L{\'o}pez}, {Managau}, {Mann}, {Mantelet}, {Marchal}, {Marchant}, {Marconi}, {Marinoni}, {Marschalk{\'o}}, {Marshall}, {Martino}, {Marton}, {Mary}, {Matijevi{\v{c}}}, {Mazeh}, {Messina}, {Michalik}, {Millar}, {Molina}, {Molinaro}, {Moln{\'a}r}, {Montegriffo}, {Mor}, {Morbidelli}, {Morel}, {Morris},
  {Mulone}, {Muraveva}, {Musella}, {Nelemans}, {Nicastro}, {Noval}, {O'Mullane}, {Ord{\'e}novic}, {Ord{\'o}{\~n}ez-Blanco}, {Osborne}, {Pagani}, {Pagano}, {Pailler}, {Palacin}, {Palaversa}, {Panahi}, {Pawlak}, {Piersimoni}, {Pineau}, {Plachy}, {Plum}, {Poggio}, {Poujoulet}, {Pr{\v{s}}a}, {Pulone}, {Racero}, {Ragaini}, {Rambaux}, {Ramos-Lerate}, {Regibo}, {Riclet}, {Ripepi}, {Riva}, {Rivard}, {Rixon}, {Roegiers}, {Roelens}, {Romero-G{\'o}mez}, {Rowell}, {Royer}, {Ruiz-Dern}, {Sadowski}, {Sagrist{\`a} Sell{\'e}s}, {Sahlmann}, {Salgado}, {Salguero}, {Sanna}, {Santana-Ros}, {Sarasso}, {Savietto}, {Schultheis}, {Sciacca}, {Segol}, {Segovia}, {S{\'e}gransan}, {Shih}, {Siltala}, {Silva}, {Smart}, {Smith}, {Solano}, {Solitro}, {Sordo}, {Soria Nieto}, {Souchay}, {Spagna}, {Spoto}, {Stampa}, {Steele}, {Steidelm{\"u}ller}, {Stephenson}, {Stoev}, {Suess}, {Surdej}, {Szabados}, {Szegedi-Elek}, {Tapiador}, {Taris}, {Tauran}, {Taylor}, {Teixeira}, {Terrett}, {Teyssandier}, {Thuillot}, {Titarenko}, {Torra Clotet}, {Turon},
  {Ulla}, {Utrilla}, {Uzzi}, {Vaillant}, {Valentini}, {Valette}, {van Elteren}, {Van Hemelryck}, {van Leeuwen}, {Vaschetto}, {Vecchiato}, {Viala}, {Vicente}, {Vogt}, {von Essen}, {Voss}, {Votruba}, {Voutsinas}, {Walmsley}, {Weiler}, {Wertz}, {Wevems}, {Wyrzykowski}, {Yoldas}, {{\v{Z}}erjal}, {Ziaeepour}, {Zorec}, {Zschocke}, {Zucker}, {Zurbach}, \& {Zwitter}}]{2018A&A...616A..12G}
{Gaia Collaboration}, {Helmi}, A., {van Leeuwen}, F., {et~al.} 2018, \aap, 616, A12, \dodoi{10.1051/0004-6361/201832698}

\bibitem[{{Gaia Collaboration} {et~al.}(2021){Gaia Collaboration}, {Brown}, {Vallenari}, {Prusti}, {de Bruijne}, {Babusiaux}, {Biermann}, {Creevey}, {Evans}, {Eyer}, {Hutton}, {Jansen}, {Jordi}, {Klioner}, {Lammers}, {Lindegren}, {Luri}, {Mignard}, {Panem}, {Pourbaix}, {Randich}, {Sartoretti}, {Soubiran}, {Walton}, {Arenou}, {Bailer-Jones}, {Bastian}, {Cropper}, {Drimmel}, {Katz}, {Lattanzi}, {van Leeuwen}, {Bakker}, {Cacciari}, {Casta{\~n}eda}, {De Angeli}, {Ducourant}, {Fabricius}, {Fouesneau}, {Fr{\'e}mat}, {Guerra}, {Guerrier}, {Guiraud}, {Jean-Antoine Piccolo}, {Masana}, {Messineo}, {Mowlavi}, {Nicolas}, {Nienartowicz}, {Pailler}, {Panuzzo}, {Riclet}, {Roux}, {Seabroke}, {Sordo}, {Tanga}, {Th{\'e}venin}, {Gracia-Abril}, {Portell}, {Teyssier}, {Altmann}, {Andrae}, {Bellas-Velidis}, {Benson}, {Berthier}, {Blomme}, {Brugaletta}, {Burgess}, {Busso}, {Carry}, {Cellino}, {Cheek}, {Clementini}, {Damerdji}, {Davidson}, {Delchambre}, {Dell'Oro}, {Fern{\'a}ndez-Hern{\'a}ndez}, {Galluccio}, {Garc{\'\i}a-Lario},
  {Garcia-Reinaldos}, {Gonz{\'a}lez-N{\'u}{\~n}ez}, {Gosset}, {Haigron}, {Halbwachs}, {Hambly}, {Harrison}, {Hatzidimitriou}, {Heiter}, {Hern{\'a}ndez}, {Hestroffer}, {Hodgkin}, {Holl}, {Jan{\ss}en}, {Jevardat de Fombelle}, {Jordan}, {Krone-Martins}, {Lanzafame}, {L{\"o}ffler}, {Lorca}, {Manteiga}, {Marchal}, {Marrese}, {Moitinho}, {Mora}, {Muinonen}, {Osborne}, {Pancino}, {Pauwels}, {Petit}, {Recio-Blanco}, {Richards}, {Riello}, {Rimoldini}, {Robin}, {Roegiers}, {Rybizki}, {Sarro}, {Siopis}, {Smith}, {Sozzetti}, {Ulla}, {Utrilla}, {van Leeuwen}, {van Reeven}, {Abbas}, {Abreu Aramburu}, {Accart}, {Aerts}, {Aguado}, {Ajaj}, {Altavilla}, {{\'A}lvarez}, {{\'A}lvarez Cid-Fuentes}, {Alves}, {Anderson}, {Anglada Varela}, {Antoja}, {Audard}, {Baines}, {Baker}, {Balaguer-N{\'u}{\~n}ez}, {Balbinot}, {Balog}, {Barache}, {Barbato}, {Barros}, {Barstow}, {Bartolom{\'e}}, {Bassilana}, {Bauchet}, {Baudesson-Stella}, {Becciani}, {Bellazzini}, {Bernet}, {Bertone}, {Bianchi}, {Blanco-Cuaresma}, {Boch}, {Bombrun}, {Bossini},
  {Bouquillon}, {Bragaglia}, {Bramante}, {Breedt}, {Bressan}, {Brouillet}, {Bucciarelli}, {Burlacu}, {Busonero}, {Butkevich}, {Buzzi}, {Caffau}, {Cancelliere}, {C{\'a}novas}, {Cantat-Gaudin}, {Carballo}, {Carlucci}, {Carnerero}, {Carrasco}, {Casamiquela}, {Castellani}, {Castro-Ginard}, {Castro Sampol}, {Chaoul}, {Charlot}, {Chemin}, {Chiavassa}, {Cioni}, {Comoretto}, {Cooper}, {Cornez}, {Cowell}, {Crifo}, {Crosta}, {Crowley}, {Dafonte}, {Dapergolas}, {David}, {David}, {de Laverny}, {De Luise}, {De March}, {De Ridder}, {de Souza}, {de Teodoro}, {de Torres}, {del Peloso}, {del Pozo}, {Delbo}, {Delgado}, {Delgado}, {Delisle}, {Di Matteo}, {Diakite}, {Diener}, {Distefano}, {Dolding}, {Eappachen}, {Edvardsson}, {Enke}, {Esquej}, {Fabre}, {Fabrizio}, {Faigler}, {Fedorets}, {Fernique}, {Fienga}, {Figueras}, {Fouron}, {Fragkoudi}, {Fraile}, {Franke}, {Gai}, {Garabato}, {Garcia-Gutierrez}, {Garc{\'\i}a-Torres}, {Garofalo}, {Gavras}, {Gerlach}, {Geyer}, {Giacobbe}, {Gilmore}, {Girona}, {Giuffrida}, {Gomel}, {Gomez},
  {Gonzalez-Santamaria}, {Gonz{\'a}lez-Vidal}, {Granvik}, {Guti{\'e}rrez-S{\'a}nchez}, {Guy}, {Hauser}, {Haywood}, {Helmi}, {Hidalgo}, {Hilger}, {H{\l}adczuk}, {Hobbs}, {Holland}, {Huckle}, {Jasniewicz}, {Jonker}, {Juaristi Campillo}, {Julbe}, {Karbevska}, {Kervella}, {Khanna}, {Kochoska}, {Kontizas}, {Kordopatis}, {Korn}, {Kostrzewa-Rutkowska}, {Kruszy{\'n}ska}, {Lambert}, {Lanza}, {Lasne}, {Le Campion}, {Le Fustec}, {Lebreton}, {Lebzelter}, {Leccia}, {Leclerc}, {Lecoeur-Taibi}, {Liao}, {Licata}, {Lindstr{\o}m}, {Lister}, {Livanou}, {Lobel}, {Madrero Pardo}, {Managau}, {Mann}, {Marchant}, {Marconi}, {Marcos Santos}, {Marinoni}, {Marocco}, {Marshall}, {Martin Polo}, {Mart{\'\i}n-Fleitas}, {Masip}, {Massari}, {Mastrobuono-Battisti}, {Mazeh}, {McMillan}, {Messina}, {Michalik}, {Millar}, {Mints}, {Molina}, {Molinaro}, {Moln{\'a}r}, {Montegriffo}, {Mor}, {Morbidelli}, {Morel}, {Morris}, {Mulone}, {Munoz}, {Muraveva}, {Murphy}, {Musella}, {Noval}, {Ord{\'e}novic}, {Orr{\`u}}, {Osinde}, {Pagani}, {Pagano},
  {Palaversa}, {Palicio}, {Panahi}, {Pawlak}, {Pe{\~n}alosa Esteller}, {Penttil{\"a}}, {Piersimoni}, {Pineau}, {Plachy}, {Plum}, {Poggio}, {Poretti}, {Poujoulet}, {Pr{\v{s}}a}, {Pulone}, {Racero}, {Ragaini}, {Rainer}, {Raiteri}, {Rambaux}, {Ramos}, {Ramos-Lerate}, {Re Fiorentin}, {Regibo}, {Reyl{\'e}}, {Ripepi}, {Riva}, {Rixon}, {Robichon}, {Robin}, {Roelens}, {Rohrbasser}, {Romero-G{\'o}mez}, {Rowell}, {Royer}, {Rybicki}, {Sadowski}, {Sagrist{\`a} Sell{\'e}s}, {Sahlmann}, {Salgado}, {Salguero}, {Samaras}, {Sanchez Gimenez}, {Sanna}, {Santove{\~n}a}, {Sarasso}, {Schultheis}, {Sciacca}, {Segol}, {Segovia}, {S{\'e}gransan}, {Semeux}, {Shahaf}, {Siddiqui}, {Siebert}, {Siltala}, {Slezak}, {Smart}, {Solano}, {Solitro}, {Souami}, {Souchay}, {Spagna}, {Spoto}, {Steele}, {Steidelm{\"u}ller}, {Stephenson}, {S{\"u}veges}, {Szabados}, {Szegedi-Elek}, {Taris}, {Tauran}, {Taylor}, {Teixeira}, {Thuillot}, {Tonello}, {Torra}, {Torra}, {Turon}, {Unger}, {Vaillant}, {van Dillen}, {Vanel}, {Vecchiato}, {Viala}, {Vicente},
  {Voutsinas}, {Weiler}, {Wevers}, {Wyrzykowski}, {Yoldas}, {Yvard}, {Zhao}, {Zorec}, {Zucker}, {Zurbach}, \& {Zwitter}}]{2021A&A...649A...1G}
{Gaia Collaboration}, {Brown}, A.~G.~A., {Vallenari}, A., {et~al.} 2021, \aap, 649, A1, \dodoi{10.1051/0004-6361/202039657}

\bibitem[{{Garavito-Camargo} {et~al.}(2021){Garavito-Camargo}, {Patel}, {Besla}, {Price-Whelan}, {G{\'o}mez}, {Laporte}, \& {Johnston}}]{2021ApJ...923..140G}
{Garavito-Camargo}, N., {Patel}, E., {Besla}, G., {et~al.} 2021, \apj, 923, 140, \dodoi{10.3847/1538-4357/ac2c05}

\bibitem[{{Gillet} {et~al.}(2015){Gillet}, {Ocvirk}, {Aubert}, {Knebe}, {Libeskind}, {Yepes}, {Gottl{\"o}ber}, \& {Hoffman}}]{2015ApJ...800...34G}
{Gillet}, N., {Ocvirk}, P., {Aubert}, D., {et~al.} 2015, \apj, 800, 34, \dodoi{10.1088/0004-637X/800/1/34}

\bibitem[{{Gnedin}(2003)}]{2003ApJ...582..141G}
{Gnedin}, O.~Y. 2003, \apj, 582, 141, \dodoi{10.1086/344636}

\bibitem[{{G{\'o}rski} {et~al.}(2005){G{\'o}rski}, {Hivon}, {Banday}, {Wandelt}, {Hansen}, {Reinecke}, \& {Bartelmann}}]{2005ApJ...622..759G}
{G{\'o}rski}, K.~M., {Hivon}, E., {Banday}, A.~J., {et~al.} 2005, \apj, 622, 759, \dodoi{10.1086/427976}

\bibitem[{{Gu} {et~al.}(2024){Gu}, {Guo}, {Cautun}, {Shao}, {Pei}, {Wang}, {Gao}, \& {Wang}}]{2024NatAs.tmp...26G}
{Gu}, Q., {Guo}, Q., {Cautun}, M., {et~al.} 2024, Nature Astronomy, \dodoi{10.1038/s41550-023-02192-6}

\bibitem[{{Gu} {et~al.}(2022){Gu}, {Guo}, {Zhang}, {Cautun}, {Lacey}, {Frenk}, \& {Shao}}]{2022MNRAS.514..390G}
{Gu}, Q., {Guo}, Q., {Zhang}, T., {et~al.} 2022, \mnras, 514, 390, \dodoi{10.1093/mnras/stac1292}

\bibitem[{{Heesters} {et~al.}(2021){Heesters}, {Habas}, {Marleau}, {M{\"u}ller}, {Duc}, {Poulain}, {Durrell}, {S{\'a}nchez-Janssen}, \& {Paudel}}]{2021A&A...654A.161H}
{Heesters}, N., {Habas}, R., {Marleau}, F.~R., {et~al.} 2021, \aap, 654, A161, \dodoi{10.1051/0004-6361/202141184}

\bibitem[{{Ibata} {et~al.}(2014{\natexlab{a}}){Ibata}, {Ibata}, {Famaey}, \& {Lewis}}]{2014Natur.511..563I}
{Ibata}, N.~G., {Ibata}, R.~A., {Famaey}, B., \& {Lewis}, G.~F. 2014{\natexlab{a}}, \nat, 511, 563, \dodoi{10.1038/nature13481}

\bibitem[{{Ibata} {et~al.}(2015){Ibata}, {Famaey}, {Lewis}, {Ibata}, \& {Martin}}]{2015ApJ...805...67I}
{Ibata}, R.~A., {Famaey}, B., {Lewis}, G.~F., {Ibata}, N.~G., \& {Martin}, N. 2015, \apj, 805, 67, \dodoi{10.1088/0004-637X/805/1/67}

\bibitem[{{Ibata} {et~al.}(2014{\natexlab{b}}){Ibata}, {Ibata}, {Lewis}, {Martin}, {Conn}, {Elahi}, {Arias}, \& {Fernando}}]{2014ApJ...784L...6I}
{Ibata}, R.~A., {Ibata}, N.~G., {Lewis}, G.~F., {et~al.} 2014{\natexlab{b}}, \apjl, 784, L6, \dodoi{10.1088/2041-8205/784/1/L6}

\bibitem[{{Ibata} {et~al.}(2013){Ibata}, {Lewis}, {Conn}, {Irwin}, {McConnachie}, {Chapman}, {Collins}, {Fardal}, {Ferguson}, {Ibata}, {Mackey}, {Martin}, {Navarro}, {Rich}, {Valls-Gabaud}, \& {Widrow}}]{2013Natur.493...62I}
{Ibata}, R.~A., {Lewis}, G.~F., {Conn}, A.~R., {et~al.} 2013, \nat, 493, 62, \dodoi{10.1038/nature11717}

\bibitem[{{Kang} {et~al.}(2005){Kang}, {Mao}, {Gao}, \& {Jing}}]{2005A&A...437..383K}
{Kang}, X., {Mao}, S., {Gao}, L., \& {Jing}, Y.~P. 2005, \aap, 437, 383, \dodoi{10.1051/0004-6361:20052675}

\bibitem[{{Kroupa} {et~al.}(2005){Kroupa}, {Theis}, \& {Boily}}]{2005A&A...431..517K}
{Kroupa}, P., {Theis}, C., \& {Boily}, C.~M. 2005, \aap, 431, 517, \dodoi{10.1051/0004-6361:20041122}

\bibitem[{{Kroupa} {et~al.}(2010){Kroupa}, {Famaey}, {de Boer}, {Dabringhausen}, {Pawlowski}, {Boily}, {Jerjen}, {Forbes}, {Hensler}, \& {Metz}}]{2010A&A...523A..32K}
{Kroupa}, P., {Famaey}, B., {de Boer}, K.~S., {et~al.} 2010, \aap, 523, A32, \dodoi{10.1051/0004-6361/201014892}

\bibitem[{{Kunkel} \& {Demers}(1976)}]{1976RGOB..182..241K}
{Kunkel}, W.~E., \& {Demers}, S. 1976, in The Galaxy and the Local Group, Vol. 182, 241

\bibitem[{{Li} {et~al.}(2021){Li}, {Hammer}, {Babusiaux}, {Pawlowski}, {Yang}, {Arenou}, {Du}, \& {Wang}}]{2021ApJ...916....8L}
{Li}, H., {Hammer}, F., {Babusiaux}, C., {et~al.} 2021, \apj, 916, 8, \dodoi{10.3847/1538-4357/ac0436}

\bibitem[{{Li} \& {Helmi}(2008)}]{2008MNRAS.385.1365L}
{Li}, Y.-S., \& {Helmi}, A. 2008, \mnras, 385, 1365, \dodoi{10.1111/j.1365-2966.2008.12854.x}

\bibitem[{{Libeskind} {et~al.}(2007){Libeskind}, {Cole}, {Frenk}, {Okamoto}, \& {Jenkins}}]{2007MNRAS.374...16L}
{Libeskind}, N.~I., {Cole}, S., {Frenk}, C.~S., {Okamoto}, T., \& {Jenkins}, A. 2007, \mnras, 374, 16, \dodoi{10.1111/j.1365-2966.2006.11205.x}

\bibitem[{{Libeskind} {et~al.}(2005){Libeskind}, {Frenk}, {Cole}, {Helly}, {Jenkins}, {Navarro}, \& {Power}}]{2005MNRAS.363..146L}
{Libeskind}, N.~I., {Frenk}, C.~S., {Cole}, S., {et~al.} 2005, \mnras, 363, 146, \dodoi{10.1111/j.1365-2966.2005.09425.x}

\bibitem[{{Libeskind} {et~al.}(2009){Libeskind}, {Frenk}, {Cole}, {Jenkins}, \& {Helly}}]{2009MNRAS.399..550L}
{Libeskind}, N.~I., {Frenk}, C.~S., {Cole}, S., {Jenkins}, A., \& {Helly}, J.~C. 2009, \mnras, 399, 550, \dodoi{10.1111/j.1365-2966.2009.15315.x}

\bibitem[{{Liu} {et~al.}(2024){Liu}, {Wang}, {Guo}, {Springel}, {Bose}, {Pakmor}, \& {Hernquist}}]{2024MNRAS.529.1405L}
{Liu}, Y., {Wang}, P., {Guo}, H., {et~al.} 2024, \mnras, 529, 1405, \dodoi{10.1093/mnras/stae625}

\bibitem[{{Lovell} {et~al.}(2011){Lovell}, {Eke}, {Frenk}, \& {Jenkins}}]{2011MNRAS.413.3013L}
{Lovell}, M.~R., {Eke}, V.~R., {Frenk}, C.~S., \& {Jenkins}, A. 2011, \mnras, 413, 3013, \dodoi{10.1111/j.1365-2966.2011.18377.x}

\bibitem[{{Lynden-Bell}(1976)}]{1976MNRAS.174..695L}
{Lynden-Bell}, D. 1976, \mnras, 174, 695, \dodoi{10.1093/mnras/174.3.695}

\bibitem[{{Mart{\'\i}nez-Delgado} {et~al.}(2021){Mart{\'\i}nez-Delgado}, {Makarov}, {Javanmardi}, {Pawlowski}, {Makarova}, {Donatiello}, {Lang}, {Rom{\'a}n}, {Vivas}, \& {Carballo-Bello}}]{2021A&A...652A..48M}
{Mart{\'\i}nez-Delgado}, D., {Makarov}, D., {Javanmardi}, B., {et~al.} 2021, \aap, 652, A48, \dodoi{10.1051/0004-6361/202141242}

\bibitem[{{McConnachie}(2012)}]{2012AJ....144....4M}
{McConnachie}, A.~W. 2012, \aj, 144, 4, \dodoi{10.1088/0004-6256/144/1/4}

\bibitem[{{Metz} {et~al.}(2007){Metz}, {Kroupa}, \& {Jerjen}}]{2007MNRAS.374.1125M}
{Metz}, M., {Kroupa}, P., \& {Jerjen}, H. 2007, \mnras, 374, 1125, \dodoi{10.1111/j.1365-2966.2006.11228.x}

\bibitem[{{Metz} {et~al.}(2009{\natexlab{a}}){Metz}, {Kroupa}, \& {Jerjen}}]{2009MNRAS.394.2223M}
---. 2009{\natexlab{a}}, \mnras, 394, 2223, \dodoi{10.1111/j.1365-2966.2009.14489.x}

\bibitem[{{Metz} {et~al.}(2008){Metz}, {Kroupa}, \& {Libeskind}}]{2008ApJ...680..287M}
{Metz}, M., {Kroupa}, P., \& {Libeskind}, N.~I. 2008, \apj, 680, 287, \dodoi{10.1086/587833}

\bibitem[{{Metz} {et~al.}(2009{\natexlab{b}}){Metz}, {Kroupa}, {Theis}, {Hensler}, \& {Jerjen}}]{2009ApJ...697..269M}
{Metz}, M., {Kroupa}, P., {Theis}, C., {Hensler}, G., \& {Jerjen}, H. 2009{\natexlab{b}}, \apj, 697, 269, \dodoi{10.1088/0004-637X/697/1/269}

\bibitem[{{Moreno} {et~al.}(2013){Moreno}, {Bluck}, {Ellison}, {Patton}, {Torrey}, \& {Moster}}]{2013MNRAS.436.1765M}
{Moreno}, J., {Bluck}, A. F.~L., {Ellison}, S.~L., {et~al.} 2013, \mnras, 436, 1765, \dodoi{10.1093/mnras/stt1694}

\bibitem[{{M{\"u}ller} {et~al.}(2015){M{\"u}ller}, {Jerjen}, \& {Binggeli}}]{2015A&A...583A..79M}
{M{\"u}ller}, O., {Jerjen}, H., \& {Binggeli}, B. 2015, \aap, 583, A79, \dodoi{10.1051/0004-6361/201526748}

\bibitem[{{M{\"u}ller} {et~al.}(2016){M{\"u}ller}, {Jerjen}, {Pawlowski}, \& {Binggeli}}]{2016A&A...595A.119M}
{M{\"u}ller}, O., {Jerjen}, H., {Pawlowski}, M.~S., \& {Binggeli}, B. 2016, \aap, 595, A119, \dodoi{10.1051/0004-6361/201629298}

\bibitem[{{M{\"u}ller} {et~al.}(2018){M{\"u}ller}, {Pawlowski}, {Jerjen}, \& {Lelli}}]{2018Sci...359..534M}
{M{\"u}ller}, O., {Pawlowski}, M.~S., {Jerjen}, H., \& {Lelli}, F. 2018, Science, 359, 534, \dodoi{10.1126/science.aao1858}

\bibitem[{{M{\"u}ller} {et~al.}(2019){M{\"u}ller}, {Rejkuba}, {Pawlowski}, {Ibata}, {Lelli}, {Hilker}, \& {Jerjen}}]{2019A&A...629A..18M}
{M{\"u}ller}, O., {Rejkuba}, M., {Pawlowski}, M.~S., {et~al.} 2019, \aap, 629, A18, \dodoi{10.1051/0004-6361/201935807}

\bibitem[{{M{\"u}ller} {et~al.}(2017){M{\"u}ller}, {Scalera}, {Binggeli}, \& {Jerjen}}]{2017A&A...602A.119M}
{M{\"u}ller}, O., {Scalera}, R., {Binggeli}, B., \& {Jerjen}, H. 2017, \aap, 602, A119, \dodoi{10.1051/0004-6361/201730434}

\bibitem[{{Mutlu-Pakdil} {et~al.}(2024){Mutlu-Pakdil}, {Sand}, {Crnojevi{\'c}}, {Bennet}, {Jones}, {Spekkens}, {Karunakaran}, {Zaritsky}, {Caldwell}, {Fielder}, {Guhathakurta}, {Seth}, {Simon}, {Strader}, \& {Toloba}}]{2024arXiv240114457M}
{Mutlu-Pakdil}, B., {Sand}, D.~J., {Crnojevi{\'c}}, D., {et~al.} 2024, arXiv e-prints, arXiv:2401.14457, \dodoi{10.48550/arXiv.2401.14457}

\bibitem[{{Nelson} {et~al.}(2019{\natexlab{a}}){Nelson}, {Springel}, {Pillepich}, {Rodriguez-Gomez}, {Torrey}, {Genel}, {Vogelsberger}, {Pakmor}, {Marinacci}, {Weinberger}, {Kelley}, {Lovell}, {Diemer}, \& {Hernquist}}]{2019ComAC...6....2N}
{Nelson}, D., {Springel}, V., {Pillepich}, A., {et~al.} 2019{\natexlab{a}}, Computational Astrophysics and Cosmology, 6, 2, \dodoi{10.1186/s40668-019-0028-x}

\bibitem[{{Nelson} {et~al.}(2019{\natexlab{b}}){Nelson}, {Pillepich}, {Springel}, {Pakmor}, {Weinberger}, {Genel}, {Torrey}, {Vogelsberger}, {Marinacci}, \& {Hernquist}}]{2019MNRAS.490.3234N}
{Nelson}, D., {Pillepich}, A., {Springel}, V., {et~al.} 2019{\natexlab{b}}, \mnras, 490, 3234, \dodoi{10.1093/mnras/stz2306}

\bibitem[{{Paudel} {et~al.}(2021){Paudel}, {Yoon}, \& {Smith}}]{2021ApJ...917L..18P}
{Paudel}, S., {Yoon}, S.-J., \& {Smith}, R. 2021, \apjl, 917, L18, \dodoi{10.3847/2041-8213/ac1866}

\bibitem[{{Pawlowski}(2018)}]{2018MPLA...3330004P}
{Pawlowski}, M.~S. 2018, Modern Physics Letters A, 33, 1830004, \dodoi{10.1142/S0217732318300045}

\bibitem[{{Pawlowski}(2021{\natexlab{a}})}]{2021Galax...9...66P}
---. 2021{\natexlab{a}}, Galaxies, 9, 66, \dodoi{10.3390/galaxies9030066}

\bibitem[{{Pawlowski}(2021{\natexlab{b}})}]{2021NatAs...5.1185P}
---. 2021{\natexlab{b}}, Nature Astronomy, 5, 1185, \dodoi{10.1038/s41550-021-01452-7}

\bibitem[{{Pawlowski} {et~al.}(2019){Pawlowski}, {Bullock}, {Kelley}, \& {Famaey}}]{2019ApJ...875..105P}
{Pawlowski}, M.~S., {Bullock}, J.~S., {Kelley}, T., \& {Famaey}, B. 2019, \apj, 875, 105, \dodoi{10.3847/1538-4357/ab10e0}

\bibitem[{{Pawlowski} {et~al.}(2015{\natexlab{a}}){Pawlowski}, {Famaey}, {Merritt}, \& {Kroupa}}]{2015ApJ...815...19P}
{Pawlowski}, M.~S., {Famaey}, B., {Merritt}, D., \& {Kroupa}, P. 2015{\natexlab{a}}, \apj, 815, 19, \dodoi{10.1088/0004-637X/815/1/19}

\bibitem[{{Pawlowski} {et~al.}(2017{\natexlab{a}}){Pawlowski}, {Ibata}, \& {Bullock}}]{2017ApJ...850..132P}
{Pawlowski}, M.~S., {Ibata}, R.~A., \& {Bullock}, J.~S. 2017{\natexlab{a}}, \apj, 850, 132, \dodoi{10.3847/1538-4357/aa9435}

\bibitem[{{Pawlowski} \& {Kroupa}(2013)}]{2013MNRAS.435.2116P}
{Pawlowski}, M.~S., \& {Kroupa}, P. 2013, \mnras, 435, 2116, \dodoi{10.1093/mnras/stt1429}

\bibitem[{{Pawlowski} \& {Kroupa}(2014)}]{2014ApJ...790...74P}
---. 2014, \apj, 790, 74, \dodoi{10.1088/0004-637X/790/1/74}

\bibitem[{{Pawlowski} \& {Kroupa}(2020)}]{2020MNRAS.491.3042P}
---. 2020, \mnras, 491, 3042, \dodoi{10.1093/mnras/stz3163}

\bibitem[{{Pawlowski} {et~al.}(2012{\natexlab{a}}){Pawlowski}, {Kroupa}, {Angus}, {de Boer}, {Famaey}, \& {Hensler}}]{2012MNRAS.424...80P}
{Pawlowski}, M.~S., {Kroupa}, P., {Angus}, G., {et~al.} 2012{\natexlab{a}}, \mnras, 424, 80, \dodoi{10.1111/j.1365-2966.2012.21169.x}

\bibitem[{{Pawlowski} {et~al.}(2011){Pawlowski}, {Kroupa}, \& {de Boer}}]{2011A&A...532A.118P}
{Pawlowski}, M.~S., {Kroupa}, P., \& {de Boer}, K.~S. 2011, \aap, 532, A118, \dodoi{10.1051/0004-6361/201015021}

\bibitem[{{Pawlowski} {et~al.}(2013){Pawlowski}, {Kroupa}, \& {Jerjen}}]{2013MNRAS.435.1928P}
{Pawlowski}, M.~S., {Kroupa}, P., \& {Jerjen}, H. 2013, \mnras, 435, 1928, \dodoi{10.1093/mnras/stt1384}

\bibitem[{{Pawlowski} \& {McGaugh}(2014)}]{2014MNRAS.440..908P}
{Pawlowski}, M.~S., \& {McGaugh}, S.~S. 2014, \mnras, 440, 908, \dodoi{10.1093/mnras/stu321}

\bibitem[{{Pawlowski} {et~al.}(2015{\natexlab{b}}){Pawlowski}, {McGaugh}, \& {Jerjen}}]{2015MNRAS.453.1047P}
{Pawlowski}, M.~S., {McGaugh}, S.~S., \& {Jerjen}, H. 2015{\natexlab{b}}, \mnras, 453, 1047, \dodoi{10.1093/mnras/stv1588}

\bibitem[{{Pawlowski} {et~al.}(2022){Pawlowski}, {Oria}, {Taibi}, {Famaey}, \& {Ibata}}]{2022ApJ...932...70P}
{Pawlowski}, M.~S., {Oria}, P.-A., {Taibi}, S., {Famaey}, B., \& {Ibata}, R. 2022, \apj, 932, 70, \dodoi{10.3847/1538-4357/ac6ce0}

\bibitem[{{Pawlowski} {et~al.}(2012{\natexlab{b}}){Pawlowski}, {Pflamm-Altenburg}, \& {Kroupa}}]{2012MNRAS.423.1109P}
{Pawlowski}, M.~S., {Pflamm-Altenburg}, J., \& {Kroupa}, P. 2012{\natexlab{b}}, \mnras, 423, 1109, \dodoi{10.1111/j.1365-2966.2012.20937.x}

\bibitem[{{Pawlowski} \& {Tony Sohn}(2021)}]{2021ApJ...923...42P}
{Pawlowski}, M.~S., \& {Tony Sohn}, S. 2021, \apj, 923, 42, \dodoi{10.3847/1538-4357/ac2aa9}

\bibitem[{{Pawlowski} {et~al.}(2014){Pawlowski}, {Famaey}, {Jerjen}, {Merritt}, {Kroupa}, {Dabringhausen}, {L{\"u}ghausen}, {Forbes}, {Hensler}, {Hammer}, {Puech}, {Fouquet}, {Flores}, \& {Yang}}]{2014MNRAS.442.2362P}
{Pawlowski}, M.~S., {Famaey}, B., {Jerjen}, H., {et~al.} 2014, \mnras, 442, 2362, \dodoi{10.1093/mnras/stu1005}

\bibitem[{{Pawlowski} {et~al.}(2017{\natexlab{b}}){Pawlowski}, {Dabringhausen}, {Famaey}, {Flores}, {Hammer}, {Hensler}, {Ibata}, {Kroupa}, {Lewis}, {Libeskind}, {McGaugh}, {Merritt}, {Puech}, \& {Yang}}]{2017AN....338..854P}
{Pawlowski}, M.~S., {Dabringhausen}, J., {Famaey}, B., {et~al.} 2017{\natexlab{b}}, Astronomische Nachrichten, 338, 854, \dodoi{10.1002/asna.201713366}

\bibitem[{{Phillips} {et~al.}(2015){Phillips}, {Cooper}, {Bullock}, \& {Boylan-Kolchin}}]{2015MNRAS.453.3839P}
{Phillips}, J.~I., {Cooper}, M.~C., {Bullock}, J.~S., \& {Boylan-Kolchin}, M. 2015, \mnras, 453, 3839, \dodoi{10.1093/mnras/stv1770}

\bibitem[{{Pillepich} {et~al.}(2018){Pillepich}, {Nelson}, {Hernquist}, {Springel}, {Pakmor}, {Torrey}, {Weinberger}, {Genel}, {Naiman}, {Marinacci}, \& {Vogelsberger}}]{2018MNRAS.475..648P}
{Pillepich}, A., {Nelson}, D., {Hernquist}, L., {et~al.} 2018, \mnras, 475, 648, \dodoi{10.1093/mnras/stx3112}

\bibitem[{{Pillepich} {et~al.}(2019){Pillepich}, {Nelson}, {Springel}, {Pakmor}, {Torrey}, {Weinberger}, {Vogelsberger}, {Marinacci}, {Genel}, {van der Wel}, \& {Hernquist}}]{2019MNRAS.490.3196P}
{Pillepich}, A., {Nelson}, D., {Springel}, V., {et~al.} 2019, \mnras, 490, 3196, \dodoi{10.1093/mnras/stz2338}

\bibitem[{{Planck Collaboration} {et~al.}(2016){Planck Collaboration}, {Ade}, {Aghanim}, {Arnaud}, {Ashdown}, {Aumont}, {Baccigalupi}, {Banday}, {Barreiro}, {Bartlett}, {Bartolo}, {Battaner}, {Battye}, {Benabed}, {Beno{\^\i}t}, {Benoit-L{\'e}vy}, {Bernard}, {Bersanelli}, {Bielewicz}, {Bock}, {Bonaldi}, {Bonavera}, {Bond}, {Borrill}, {Bouchet}, {Boulanger}, {Bucher}, {Burigana}, {Butler}, {Calabrese}, {Cardoso}, {Catalano}, {Challinor}, {Chamballu}, {Chary}, {Chiang}, {Chluba}, {Christensen}, {Church}, {Clements}, {Colombi}, {Colombo}, {Combet}, {Coulais}, {Crill}, {Curto}, {Cuttaia}, {Danese}, {Davies}, {Davis}, {de Bernardis}, {de Rosa}, {de Zotti}, {Delabrouille}, {D{\'e}sert}, {Di Valentino}, {Dickinson}, {Diego}, {Dolag}, {Dole}, {Donzelli}, {Dor{\'e}}, {Douspis}, {Ducout}, {Dunkley}, {Dupac}, {Efstathiou}, {Elsner}, {En{\ss}lin}, {Eriksen}, {Farhang}, {Fergusson}, {Finelli}, {Forni}, {Frailis}, {Fraisse}, {Franceschi}, {Frejsel}, {Galeotta}, {Galli}, {Ganga}, {Gauthier}, {Gerbino}, {Ghosh}, {Giard},
  {Giraud-H{\'e}raud}, {Giusarma}, {Gjerl{\o}w}, {Gonz{\'a}lez-Nuevo}, {G{\'o}rski}, {Gratton}, {Gregorio}, {Gruppuso}, {Gudmundsson}, {Hamann}, {Hansen}, {Hanson}, {Harrison}, {Helou}, {Henrot-Versill{\'e}}, {Hern{\'a}ndez-Monteagudo}, {Herranz}, {Hildebrandt}, {Hivon}, {Hobson}, {Holmes}, {Hornstrup}, {Hovest}, {Huang}, {Huffenberger}, {Hurier}, {Jaffe}, {Jaffe}, {Jones}, {Juvela}, {Keih{\"a}nen}, {Keskitalo}, {Kisner}, {Kneissl}, {Knoche}, {Knox}, {Kunz}, {Kurki-Suonio}, {Lagache}, {L{\"a}hteenm{\"a}ki}, {Lamarre}, {Lasenby}, {Lattanzi}, {Lawrence}, {Leahy}, {Leonardi}, {Lesgourgues}, {Levrier}, {Lewis}, {Liguori}, {Lilje}, {Linden-V{\o}rnle}, {L{\'o}pez-Caniego}, {Lubin}, {Mac{\'\i}as-P{\'e}rez}, {Maggio}, {Maino}, {Mandolesi}, {Mangilli}, {Marchini}, {Maris}, {Martin}, {Martinelli}, {Mart{\'\i}nez-Gonz{\'a}lez}, {Masi}, {Matarrese}, {McGehee}, {Meinhold}, {Melchiorri}, {Melin}, {Mendes}, {Mennella}, {Migliaccio}, {Millea}, {Mitra}, {Miville-Desch{\^e}nes}, {Moneti}, {Montier}, {Morgante}, {Mortlock},
  {Moss}, {Munshi}, {Murphy}, {Naselsky}, {Nati}, {Natoli}, {Netterfield}, {N{\o}rgaard-Nielsen}, {Noviello}, {Novikov}, {Novikov}, {Oxborrow}, {Paci}, {Pagano}, {Pajot}, {Paladini}, {Paoletti}, {Partridge}, {Pasian}, {Patanchon}, {Pearson}, {Perdereau}, {Perotto}, {Perrotta}, {Pettorino}, {Piacentini}, {Piat}, {Pierpaoli}, {Pietrobon}, {Plaszczynski}, {Pointecouteau}, {Polenta}, {Popa}, {Pratt}, {Pr{\'e}zeau}, {Prunet}, {Puget}, {Rachen}, {Reach}, {Rebolo}, {Reinecke}, {Remazeilles}, {Renault}, {Renzi}, {Ristorcelli}, {Rocha}, {Rosset}, {Rossetti}, {Roudier}, {Rouill{\'e} d'Orfeuil}, {Rowan-Robinson}, {Rubi{\~n}o-Mart{\'\i}n}, {Rusholme}, {Said}, {Salvatelli}, {Salvati}, {Sandri}, {Santos}, {Savelainen}, {Savini}, {Scott}, {Seiffert}, {Serra}, {Shellard}, {Spencer}, {Spinelli}, {Stolyarov}, {Stompor}, {Sudiwala}, {Sunyaev}, {Sutton}, {Suur-Uski}, {Sygnet}, {Tauber}, {Terenzi}, {Toffolatti}, {Tomasi}, {Tristram}, {Trombetti}, {Tucci}, {Tuovinen}, {T{\"u}rler}, {Umana}, {Valenziano}, {Valiviita}, {Van Tent},
  {Vielva}, {Villa}, {Wade}, {Wandelt}, {Wehus}, {White}, {White}, {Wilkinson}, {Yvon}, {Zacchei}, \& {Zonca}}]{2016A&A...594A..13P}
{Planck Collaboration}, {Ade}, P.~A.~R., {Aghanim}, N., {et~al.} 2016, \aap, 594, A13, \dodoi{10.1051/0004-6361/201525830}

\bibitem[{{Samuel} {et~al.}(2021){Samuel}, {Wetzel}, {Chapman}, {Tollerud}, {Hopkins}, {Boylan-Kolchin}, {Bailin}, \& {Faucher-Gigu{\`e}re}}]{2021MNRAS.504.1379S}
{Samuel}, J., {Wetzel}, A., {Chapman}, S., {et~al.} 2021, \mnras, 504, 1379, \dodoi{10.1093/mnras/stab955}

\bibitem[{{Santos-Santos} {et~al.}(2020{\natexlab{a}}){Santos-Santos}, {Dom{\'\i}nguez-Tenreiro}, {Artal}, {Pedrosa}, {Bignone}, {Mart{\'\i}nez-Serrano}, {G{\'o}mez-Flechoso}, {Tissera}, \& {Serna}}]{2020ApJ...897...71S}
{Santos-Santos}, I., {Dom{\'\i}nguez-Tenreiro}, R., {Artal}, H., {et~al.} 2020{\natexlab{a}}, \apj, 897, 71, \dodoi{10.3847/1538-4357/ab7f29}

\bibitem[{{Santos-Santos} {et~al.}(2020{\natexlab{b}}){Santos-Santos}, {Dom{\'\i}nguez-Tenreiro}, \& {Pawlowski}}]{2020MNRAS.499.3755S}
{Santos-Santos}, I.~M., {Dom{\'\i}nguez-Tenreiro}, R., \& {Pawlowski}, M.~S. 2020{\natexlab{b}}, \mnras, 499, 3755, \dodoi{10.1093/mnras/staa3130}

\bibitem[{{Santos-Santos} {et~al.}(2021){Santos-Santos}, {Fattahi}, {Sales}, \& {Navarro}}]{2021MNRAS.504.4551S}
{Santos-Santos}, I. M.~E., {Fattahi}, A., {Sales}, L.~V., \& {Navarro}, J.~F. 2021, \mnras, 504, 4551, \dodoi{10.1093/mnras/stab1020}

\bibitem[{{Sawala} {et~al.}(2014){Sawala}, {Frenk}, {Fattahi}, {Navarro}, {Bower}, {Crain}, {Dalla Vecchia}, {Furlong}, {Helly}, {Jenkins}, {Oman}, {Schaller}, {Schaye}, {Theuns}, {Trayford}, \& {White}}]{2014arXiv1412.2748S}
{Sawala}, T., {Frenk}, C.~S., {Fattahi}, A., {et~al.} 2014, arXiv e-prints, arXiv:1412.2748, \dodoi{10.48550/arXiv.1412.2748}

\bibitem[{{Sawala} {et~al.}(2016){Sawala}, {Frenk}, {Fattahi}, {Navarro}, {Bower}, {Crain}, {Dalla Vecchia}, {Furlong}, {Helly}, {Jenkins}, {Oman}, {Schaller}, {Schaye}, {Theuns}, {Trayford}, \& {White}}]{2016MNRAS.457.1931S}
---. 2016, \mnras, 457, 1931, \dodoi{10.1093/mnras/stw145}

\bibitem[{{Sawala} {et~al.}(2023){Sawala}, {Cautun}, {Frenk}, {Helly}, {Jasche}, {Jenkins}, {Johansson}, {Lavaux}, {McAlpine}, \& {Schaller}}]{2023NatAs...7..481S}
{Sawala}, T., {Cautun}, M., {Frenk}, C., {et~al.} 2023, Nature Astronomy, 7, 481, \dodoi{10.1038/s41550-022-01856-z}

\bibitem[{{Shao} {et~al.}(2019){Shao}, {Cautun}, \& {Frenk}}]{2019MNRAS.488.1166S}
{Shao}, S., {Cautun}, M., \& {Frenk}, C.~S. 2019, \mnras, 488, 1166, \dodoi{10.1093/mnras/stz1741}

\bibitem[{{Shao} {et~al.}(2018){Shao}, {Cautun}, {Frenk}, {Grand}, {G{\'o}mez}, {Marinacci}, \& {Simpson}}]{2018MNRAS.476.1796S}
{Shao}, S., {Cautun}, M., {Frenk}, C.~S., {et~al.} 2018, \mnras, 476, 1796, \dodoi{10.1093/mnras/sty343}

\bibitem[{{Shipp} {et~al.}(2021){Shipp}, {Erkal}, {Drlica-Wagner}, {Li}, {Pace}, {Koposov}, {Cullinane}, {Da Costa}, {Ji}, {Kuehn}, {Lewis}, {Mackey}, {Simpson}, {Wan}, {Zucker}, {Bland-Hawthorn}, {Ferguson}, {Lilleengen}, \& {Lilleengen}}]{2021ApJ...923..149S}
{Shipp}, N., {Erkal}, D., {Drlica-Wagner}, A., {et~al.} 2021, \apj, 923, 149, \dodoi{10.3847/1538-4357/ac2e93}

\bibitem[{{Sinha} \& {Holley-Bockelmann}(2012)}]{2012ApJ...751...17S}
{Sinha}, M., \& {Holley-Bockelmann}, K. 2012, \apj, 751, 17, \dodoi{10.1088/0004-637X/751/1/17}

\bibitem[{{Sohn} {et~al.}(2020){Sohn}, {Patel}, {Fardal}, {Besla}, {van der Marel}, {Geha}, \& {Guhathakurta}}]{2020ApJ...901...43S}
{Sohn}, S.~T., {Patel}, E., {Fardal}, M.~A., {et~al.} 2020, \apj, 901, 43, \dodoi{10.3847/1538-4357/abaf49}

\bibitem[{{Springel} {et~al.}(2001){Springel}, {White}, {Tormen}, \& {Kauffmann}}]{2001MNRAS.328..726S}
{Springel}, V., {White}, S. D.~M., {Tormen}, G., \& {Kauffmann}, G. 2001, \mnras, 328, 726, \dodoi{10.1046/j.1365-8711.2001.04912.x}

\bibitem[{{Springel} {et~al.}(2018){Springel}, {Pakmor}, {Pillepich}, {Weinberger}, {Nelson}, {Hernquist}, {Vogelsberger}, {Genel}, {Torrey}, {Marinacci}, \& {Naiman}}]{2018MNRAS.475..676S}
{Springel}, V., {Pakmor}, R., {Pillepich}, A., {et~al.} 2018, \mnras, 475, 676, \dodoi{10.1093/mnras/stx3304}

\bibitem[{{Stanimirovi{\'c}} {et~al.}(2004){Stanimirovi{\'c}}, {Staveley-Smith}, \& {Jones}}]{2004ApJ...604..176S}
{Stanimirovi{\'c}}, S., {Staveley-Smith}, L., \& {Jones}, P.~A. 2004, \apj, 604, 176, \dodoi{10.1086/381869}

\bibitem[{{Taibi} {et~al.}(2024){Taibi}, {Pawlowski}, {Khoperskov}, {Steinmetz}, \& {Libeskind}}]{2024A&A...681A..73T}
{Taibi}, S., {Pawlowski}, M.~S., {Khoperskov}, S., {Steinmetz}, M., \& {Libeskind}, N.~I. 2024, \aap, 681, A73, \dodoi{10.1051/0004-6361/202347473}

\bibitem[{{Tonnesen} \& {Cen}(2012)}]{2012MNRAS.425.2313T}
{Tonnesen}, S., \& {Cen}, R. 2012, \mnras, 425, 2313, \dodoi{10.1111/j.1365-2966.2012.21637.x}

\bibitem[{{Tully} {et~al.}(2015){Tully}, {Libeskind}, {Karachentsev}, {Karachentseva}, {Rizzi}, \& {Shaya}}]{2015ApJ...802L..25T}
{Tully}, R.~B., {Libeskind}, N.~I., {Karachentsev}, I.~D., {et~al.} 2015, \apjl, 802, L25, \dodoi{10.1088/2041-8205/802/2/L25}

\bibitem[{{Vasiliev}(2023)}]{2023Galax..11...59V}
{Vasiliev}, E. 2023, Galaxies, 11, 59, \dodoi{10.3390/galaxies11020059}

\bibitem[{{Wang} {et~al.}(2013){Wang}, {Frenk}, \& {Cooper}}]{2013MNRAS.429.1502W}
{Wang}, J., {Frenk}, C.~S., \& {Cooper}, A.~P. 2013, \mnras, 429, 1502, \dodoi{10.1093/mnras/sts442}

\bibitem[{{Wang} {et~al.}(2020){Wang}, {Libeskind}, {Tempel}, {Pawlowski}, {Kang}, \& {Guo}}]{2020ApJ...900..129W}
{Wang}, P., {Libeskind}, N.~I., {Tempel}, E., {et~al.} 2020, \apj, 900, 129, \dodoi{10.3847/1538-4357/aba6ea}

\bibitem[{{Welker} {et~al.}(2018){Welker}, {Dubois}, {Pichon}, {Devriendt}, \& {Chisari}}]{2018A&A...613A...4W}
{Welker}, C., {Dubois}, Y., {Pichon}, C., {Devriendt}, J., \& {Chisari}, N.~E. 2018, \aap, 613, A4, \dodoi{10.1051/0004-6361/201629007}

\bibitem[{{Xu} {et~al.}(2023){Xu}, {Kang}, \& {Libeskind}}]{2023ApJ...954..128X}
{Xu}, Y., {Kang}, X., \& {Libeskind}, N.~I. 2023, \apj, 954, 128, \dodoi{10.3847/1538-4357/ace898}

\bibitem[{{Yoon} \& {Lee}(2002)}]{2002Sci...297..578Y}
{Yoon}, S.-J., \& {Lee}, Y.-W. 2002, Science, 297, 578, \dodoi{10.1126/science.1073090}

\bibitem[{{Zentner} {et~al.}(2005){Zentner}, {Kravtsov}, {Gnedin}, \& {Klypin}}]{2005ApJ...629..219Z}
{Zentner}, A.~R., {Kravtsov}, A.~V., {Gnedin}, O.~Y., \& {Klypin}, A.~A. 2005, \apj, 629, 219, \dodoi{10.1086/431355}

\bibitem[{{Zhao} {et~al.}(2023){Zhao}, {Mathews}, {Phillips}, \& {Tang}}]{2023Galax..11..114Z}
{Zhao}, X., {Mathews}, G.~J., {Phillips}, L.~A., \& {Tang}, G. 2023, Galaxies, 11, 114, \dodoi{10.3390/galaxies11060114}

\bibitem[{Zonca {et~al.}(2019)Zonca, Singer, Lenz, Reinecke, Rosset, Hivon, \& Gorski}]{Zonca2019}
Zonca, A., Singer, L., Lenz, D., {et~al.} 2019, Journal of Open Source Software, 4, 1298, \dodoi{10.21105/joss.01298}

\end{thebibliography}

\begin{table*}
\centering
\begin{tabular}{ccccccccccc}
\hline \hline
& \multicolumn{3}{c}{Position Data} & \multicolumn{3}{c}{Velocity Data} & \multicolumn{2}{c}{Total Mass Data}\\
Name & $x$     & $y$     & $z$     & $V_{\rm x}$ & $V_{\rm y}$ & $V_{\rm z}$ & $M_{\rm tot,cusp}$ & $M_{\rm tot,core}$ \\
                 & [kpc] & [kpc] & [kpc] & [km/s] & [km/s] & [km/s] & [$log_{10} {M_{tot}\over M_{\odot}}$] & [$log_{10} {M_{tot}\over M_{\odot}}$] \\ \hline
LMC              & $-$0.6   & $-$41.8  & $-$27.5 & $-$42 $\pm6$ & $-$223 $\pm4$ & 231 $\pm4$ &  \multicolumn{2}{c}{${11.27^{+0.08}_{-0.1}}$} \\ 
SMC              & 16.5   & $-$38.5  & $-$44.7 & 6 $\pm8$ & $-$180 $\pm7$ & 167 $\pm6$ & \multicolumn{2}{c}{$>9.38$} \\ 
Sagittarius dSph & 17.1   & 2.5  & $-$6.4 & 233 $\pm2$ & $-$8 $\pm4$ & 209 $\pm4$ & ${8.8^{+0.2}_{-0.1}}$ & ${8.9^{+0.0}_{-0.1}}$ \\ 
Fornax dSph & $-$39.6   & $-$48.3  & $-$126.9 & 15 $\pm11$ & $-$101 $\pm19$ & 77 $\pm7$ & ${8.9^{+0.5}_{-0.2}}$ & ${9.4^{+1.0}_{-0.2}}$ \\ 
Leo I dSph & $-$123.2  & $-$119.6  & 192.3 & $-$122.8 $\pm5.3$ & $-$15.6 $\pm10.2$ & 137.3 $\pm6.2$  & ${9.3^{+0.3}_{-0.5}}$ & ${10.4^{+0.1}_{-0.0}}$ \\ 
Sculptor dSph & $-$5.3   & $-$9.5  & $-$83.4 & 17 $\pm8$ & 155 $\pm8$ & $-$94 $\pm1$ & ${9.3^{+0.2}_{-0.8}}$ & ${10.4^{+0.2}_{-0.7}}$ \\ 
Leo II dSph & $-$76.5  & $-$58.2 & 215.0 & $-$89.3 $\pm10.7$ & 42.6 $\pm15.5$ & 4.9 $\pm5.9$ & ${8.5^{+0.8}_{-0.4}}$ & ${10.4^{+0.2}_{-1.9}}$ \\ 
Sextans (I) dSph  & $-$35.9  & $-$56.1  & 57.1 & $-$197 $\pm10$ & 70 $\pm6$ & 73 $\pm7$ & ${8.5^{+0.3}_{-0.5}}$ & ${8.6^{+0.3}_{-0.5}}$ \\ 
Carina dSph & $-$25.1  & $-$96.5  & $-$39.9 & $-$29 $\pm10$ & $-$66 $\pm7$ & 175 $\pm17$ & ${8.6^{+0.4}_{-1.0}}$ & ${9.3^{+1.3}_{-1.2}}$ \\ 
Draco dSph  & $-$3.8  & 66.1  & 45.9 & 68 $\pm9$ & 4 $\pm4$ & $-$170 $\pm6$ & ${9.5^{+0.3}_{-0.8}}$ & ${10.4^{+0.2}_{-0.4}}$ \\ 
Ursa Minor dSph  & $-$22.0  & 52.3  & 53.8 & 10 $\pm7$ & 50 $\pm5$ & $-$157 $\pm5$ & ${8.9^{+0.9}_{-0.5}}$ & ${10.4^{+0.2}_{-0.7}}$ \\ 
\hline \hline
\end{tabular}
\caption{({\it Left: Position data}) The positions of the classical 11 satellites. Data for the LMC, SMC, Sgr dSph, Leo I, and Leo II are taken from \citet{2020MNRAS.491.3042P} and \citet{2024ApJ...971...98B}, while data for the other six satellites are from \citet{2024A&A...681A..73T}. The x-, y-, and z-values, in units of kpc, are provided in Galactocentric Cartesian coordinates. ({\it Middle: Velocity data}) The velocities of the 11 classical satellites of the Milky Way. Sources for the velocity data include \citet{2020MNRAS.491.3042P} for LMC, SMC, and Sgr dSph; \citet{2024ApJ...971...98B} for Leo I and Leo II; and \citet{2024A&A...681A..73T} for the other six satellites. ({\it Right: Total mass data}) The total halo masses of the 11 classical satellites. The LMC mass is from \citet{2021ApJ...923..149S}, the SMC mass is from \citet{2004ApJ...604..176S}, and the masses of the rest of satellites are from \citet{2018MNRAS.481.5073E}. For the nine classical satellites (excluding LMC and SMC), the total masses are estimated with two different (cusp and core) DM profiles.}
\label{tab:used data}
\end{table*}

\begin{table*}
\centering
\begin{tabular}{clc}
\hline \hline
Sample Selection & \hspace{1.5cm} Rarity of the MW DoS & Relevant Figures \\
\hline \hline
\multicolumn{3}{l}{The Conventional $c$/$a$-based Estimation Using Present-day $c$/$a$} \\
\hline
\hspace{1.425cm} Case A (Fiducial)    & \hspace{1.9cm} 1.49\,\% (= 3/202) & Figures 1, 9, 12 \\
Case B               & \hspace{1.9cm} 1.46\,\% (= 3/206) & Figure 9         \\
Case C               & \hspace{1.9cm} 0.00\,\% (= 0/202) & Figure 9         \\
Case D               & \hspace{1.9cm} 1.71\,\% (= 4/234) & Figure 12        \\
\hline \hline                
\multicolumn{3}{l}{The New `$c$/$a$ PDF'-based Estimation Using the Median of $c$/$a$ PDF ($\mu_{\rm PDF}$)} \\
\hline                 
\hspace{1.425cm} Case A (Fiducial)    & \hspace{1.9cm} 2.48\,\% (= 5/202) & Figures 5, 10, 13 \\
Case B               & \hspace{1.9cm} 3.40\,\% (= 7/206) & Figure 10         \\
Case C               & \hspace{1.9cm} 0.00\,\% (= 0/202) & Figure 10         \\
Case D               & \hspace{1.9cm} 2.99\,\% (= 7/234) & Figure 13         \\ 
\hline \hline
\multicolumn{3}{l}{The `$\langle d \rangle_{\rm norm}$ vs. $min$($\alpha_{N}$) Plane'-based Estimation Using the Distance from the TNG50-1 Systems} \\
\hline                 
\hspace{1.425cm} Case A (Fiducial)    & \hspace{1.9cm} 0.64\,\% (2.49\,$\sigma$) & Figures 8, 11, 14 \\
Case B               & \hspace{1.9cm} 0.59\,\% (2.52\,$\sigma$) & Figure 11         \\
Case C               & \hspace{1.9cm} 0.03\,\% (3.41\,$\sigma$) & Figure 11         \\
Case D               & \hspace{1.9cm} 0.76\,\% (2.43\,$\sigma$) & Figure 14         \\
\hline \hline
\multicolumn{3}{l}{The `$min$($\alpha_{N}$) vs. $N_{\rm corot}$ Plane'-based Estimation Using the Number of the TNG50-1 Systems} \\
\hline                 
\hspace{1.425cm} Case A (Fiducial)     & \hspace{1.9cm} 0.50\,\% (= 1/202) & Figure 15  \\
\hline \hline
\end{tabular}
\caption{
The summary of the estimated rarities of the MW DoS. We use four different methodologies: (1) The conventional $c$/$a$-based estimation using present-day $c$/$a$; (2) The New `$c$/$a$ PDF'-based estimation using the median of $c$/$a$ PDF ($\mu_{\rm PDF}$); (3) The `$\langle d \rangle_{\rm norm}$ vs. $min$($\alpha_{N}$) plane'-based estimation using the distance from the MW-analogous host--satellite systems in the TNG50-1; and (4) The `$min$($\alpha_{N}$) vs. $N_{\rm corot}$ plane'-based estimation using the number of the MW-analogous host--satellite systems in the TNG50-1. 
({\it Column 1}) We use four different sample selection methods for the MW-analogous host--satellite systems in the TNG50-1: (Case A) The fiducial sample, i.e., all SUBFIND halos with the total energy cut; (Case B) All SUBFIND halos with no additional cut; (Case C) All SUBFIND halos with the 1.3\,$R_{\rm vir}$ cut; and (Case D) The fiducial + massive-LMC (the total mass of the LMC is assumed to be 20\,\% of that of the MW) assumption. 
({\it Column 2}) The resultant rarity estimations for the MW DoS and the quantities used to calculate the rarities. For the `$\langle d \rangle_{\rm norm}$ vs. $min$($\alpha_{N}$) plane'-based estimation, we assume 2-D Gaussian distributions of $\langle d \rangle_{\rm norm}$ and $min$($\alpha_{N}$), which are independent of one another. 
({\it Column 3}) The relevant Figures in the text. 
}
\label{tab:raritysummary}
\end{table*}

\end{document}